\documentclass[prd,nofootinbib,aps,twocolumn,floats,floatfix,
superscriptaddress,amsmath,amssymb,secnumarabic]{revtex4}
\usepackage{natbib}
\usepackage{moreverb}
\usepackage{graphicx}
\usepackage{amsmath}
\usepackage{amssymb}
\usepackage{float}
\usepackage{slashed}
\usepackage{color}
\usepackage[utf8]{inputenc}
\usepackage{hyperref}

\newcommand{\be}{\begin{equation}}
\newcommand{\ee}{\end{equation}}
\newcommand{\ba}{\begin{array}}
\newcommand{\ea}{\end{array}}
\newcommand{\bea}{\begin{eqnarray}}
\newcommand{\eea}{\end{eqnarray}}

\def\sfrac#1#2{{\textstyle{#1\over #2}}}

\begin{document}

\title{$p$-wave Annihilating Dark Matter from a Decaying Predecessor
\\and the Galactic Center Excess}
\author{Jeremie Choquette\footnote{jeremie.choquette@physics.mcgill.ca}}
\author{James M.\ Cline\footnote{jcline@physics.mcgill.ca}}
\author{Jonathan M. Cornell\footnote{cornellj@physics.mcgill.ca}}
\affiliation{Department of Physics, McGill University,
3600 Rue University, Montr\'eal, Qu\'ebec, Canada H3A 2T8}
\begin{abstract}
Dark matter (DM) annihilations have been widely studied as a possible
explanation of excess gamma rays from the galactic center seen by
Fermi/LAT.  However most such models are in conflict with constraints from
dwarf spheroidals.  Motivated by this tension, we show that $p$-wave
annihilating dark matter can easily accommodate both sets of
observations due to the lower DM velocity dispersion in dwarf
galaxies.  Explaining the DM relic abundance is then challenging.
We outline a scenario in which the usual thermal abundance is obtained
through $s$-wave annihilations of a metastable particle, that
eventually 
decays into the $p$-wave annihilating DM of the present epoch.  The
couplings and lifetime of the decaying particle are constrained by
big bang nucleosynthesis, the cosmic microwave background and direct
detection, but
significant regions of parameter space are viable.  A sufficiently
large $p$-wave cross section can be found by annihilation into light
mediators, that also give rise to Sommerfeld enhancement.   A prediction
of the scenario is enhanced annihilations in galaxy clusters.
\end{abstract}
\maketitle

Fermi-LAT Observations of the galactic center (GC) provide evidence of
a gamma-ray excess in the multi-GeV energy range 
\cite{Goodenough:2009gk,Hooper:2010mq,Hooper:2011ti,
Abazajian:2012pn,Zhou:2014lva,Calore:2014xka,Daylan:2014rsa,FermiData,TheFermi-LAT:2015kwa}. 
Millisecond pulsars (MSPs) are a favored astrophysical source to explain the
signal
\cite{Lee:2015fea,Bartels:2015aea,O'Leary:2015gfa}, but
there is debate in the literature as to whether the required numbers
of MSPs in the GC for this purpose is consistent with the number 
resolved by Fermi \cite{Linden:2015qha} or expected on theoretical or
empirical grounds \cite{Cholis:2014lta}.  Other possible
astrophysical explanations have been presented
\cite{Carlson:2014cwa, Petrovic:2014uda, Cholis:2015dea, Gaggero:2015nsa, deBoer:2015kta,Carlson:2016iis}, but dark matter annihilation
into charged particles that lead to gamma rays remains a possibility
that has attracted great interest.   Further data should eventually be
able to distinguish between the different possibilities
\cite{Yuan:2014yda,Lee:2014mza}.

There is tension between most dark matter (DM) explanations of the
galactic center excess (GCE) and constraints on dark matter annihilation coming from
observations of dwarf spheroidal galaxies~\cite{Ackermann:2015zua,Abazajian:2015raa}.\footnote{Analyses of known dwarf spheroidal galaxies have revealed no significant excess gamma-ray emission. However, there have been claims of possible signals from the recently discovered \cite{Bechtol:2015cbp, Drlica-Wagner:2015ufc} dwarf spheroidal candidates Reticulum II \cite{Geringer-Sameth:2015lua, Hooper:2015ula} and Turcana III \cite{Li:2015kag}. These results are somewhat in dispute, with a Fermi-LAT analysis of Reticulum II using more data \cite{Drlica-Wagner:2015xua} claiming no significant excess.}
(Further complementary constraints come from searches for GeV emission
in the large Magellanic cloud \cite{Buckley:2015doa} or subhalos
of the Milky Way 
\cite{Bertoni:2015mla}.)
The best fits for DM mass and annihilation cross section for the
GCE lie in regions that tend to be excluded by factors of a 
few by the dwarf spheroidal limits.  A possible way of alleviating
this tension is to assume that the annihilation is into electrons,
a scenario in which the GCE is primarily produced through inverse Compton scattering which is suppressed in dwarfs because of their dilute
radiation fields 
\cite{Liu:2014cma,Calore:2014nla,Kaplinghat:2015gha}. An additional idea to explain the discrepancy is a model of asymmetric DM where anti-DM is produced at late times via decays, leading to particles with enough kinetic energy to escape a dwarf galaxy but not the galactic center, where they annihilate with DM particles \cite{Hardy:2014dea}.

In this work we explore a different possibility, noting that 
the tension can be avoided if the dark matter annihilation rates are
velocity-dependent. Since the velocity dispersion in the galactic
center is significantly higher than that in dwarf galaxies, the
GCE can be consistent with the lack of
signals from dwarf spheroidal galaxies provided that the annihilation
cross section increases with velocity. This is the case in models
where $p$-wave annihilations dominate, which is the subject of the
present work.
This scenario has recently been explored~\cite{Zhao:2016xie} to alleviate tension between the dwarf spheroidal constraints and DM explanations of the AMS-02 positron excess. We take a similar approach for the GCE. 
An immediate challenge is
how to obtain the right relic density since the cross section needed
for the GCE is of order $\langle\sigma v\rangle \sim 3\times 10^{26}
{\rm cm^3/s}$, the usual value associated with a thermal origin for
the relic density.
But if $\langle\sigma v\rangle$ has such a value in the GC today, it
would have been orders of magnitude larger in the early universe,
leading to a negligible thermal abundance.  We address this by
showing how the current generation of $p$-wave annihilating
dark matter could have arisen through the decays of a metastable 
predecessor DM particle that has a thermal origin.  The decays can
take place at temperatures ranging from $\sim 1$ eV to several GeV. 
By this time the $p$-wave
annihilations would be out of equilibrium despite their relatively
large cross section.

The annihilation cross section needed to explain the GCE requires 
large couplings to compensate for the $p$-wave suppresssion.  Such
large copulings would generically tend to also give strong
interactions of dark matter with nuclei.  However constraints from
direct detection can be satisfied if the dark matter annihilates into
light mediators \cite{Abdullah:2014lla,Elor:2015tva} that 
subsequently decay into standard model particles.  The light mediators also lead to Sommerfeld enhanced annihilation, allowing us to avoid nonperturbatively large couplings. In this way we are
able to find viable models that have reasonably small couplings.

In section~\ref{sec:radial} we parametrize the $p$-wave
 annihilation cross section
in the Milky Way (MW) and in
 dwarf spheroidal galaxies, in terms of assumed velocity dispersion
profiles, leading to 
modified $J$-factors that are relevant for comparison to observations.
In section~\ref{sec:simulation} we give the
results of the galactic propagation simulations used to compute the expected signal from
the galactic center, including the effects of inverse Compton
scattering and Bremsstrahlung radiation.  This yields fits to the
data in the plane of DM mass versus annihilation cross section
$\sigma v$.
We then derive upper limits on $\sigma v$ in the same plane from 
dwarf spheroidals and galaxy clusters.
In section~\ref{sec:relic}
we show that $p$-wave annihilations of the desired strength would lead
to strong suppression of the DM abundance at freeze-out, unless some 
nonthermal origin prevails. Here we present the scenario of decaying DM 
whose density is determined by the usual $s$-wave process, 
and the conditions under which this provides a consistent description.
Three examples of decay channels leading to different phenomenology
are presented,  to illustrate the range of possibilities.
In section \ref{constraints} we systematically explore observational
constraints on these models coming from cosmology, astrophysical line
searches, direct searches, and colliders.  In section \ref{mediator}
we provide a concrete model of $\chi$ annihilation into light scalar
mediators to show that the desired large cross section can be achieved
with reasonable values of the couplings in a renormalizable model.
Conclusions are given in section \ref{conclusion}.

\section{Annihilation Cross Section}
\label{sec:radial}

The expected signal from either the GC or dwarf spheroidals is
proportional to the phase-space averaged 
cross section,
\begin{align}
\langle\sigma v\rangle=\sfrac12\int_0^{v_{\rm esc}}dv_1
\int_0^{v_{\rm esc}}dv_2\int_{-1}^1d\cos\theta\,f(v_1)\,f(v_2) 
\,\sigma v_{\rm rel}
\label{eq:velavg}
\end{align}
for a velocity distribution $f(v)$, where $v_{\rm
rel}=\sqrt{v_1^2+v_2^2-2v_1v_2\cos\theta}$ is the relative velocity between the
two annihilating particles and the escape velocity $v_{\rm esc}$
depends upon radial position $r$ in the galaxy. In this work we consider
Dirac fermion dark matter. Self-annihilating Majorana dark matter would 
introduce an additional factor of $1/2$ into equation
(\ref{eq:velavg}). Following~\cite{Robertson:2009bh} and
others, we adopt a Maxwell-Boltzmann distribution,
\begin{align}
f(v)=\frac{3\sqrt{6}}{\sqrt{\pi}\sigma_v^3}\, v^2\,e^{-3v^2/2\sigma_v^2},
\label{eq:f}
\end{align} 
where $\sigma_v$ is the velocity dispersion at the given $r$.
The normalization factor in (\ref{eq:f}) is appropriate in the 
limit of large escape velocity, $v_{\rm esc}\gg \sigma_v$. Numerically
 we find that this approximation is well-suited to the present
applications.

We will be interested in $p$-wave annihilation for which at low
velocities $\sigma v \cong \frac12 C_\sigma (v/c)^2$ with $C_\sigma$ a constant.
The phase-space averaged value is then 
\be
\langle\sigma v \rangle  = C_\sigma (\sigma_v/c)^2
\label{eq:Csigma}
\ee
In general $\sigma_v$ is a function of $r$.   This dependence is
potentially significant in the MW, unlike in dwarf spheroidals, whose
radial dependence has been observed to be roughly constant. 
Regardless of these details, it is however clear that 
$\langle\sigma v \rangle$ is several orders of magnitude lower in
dwarf spheroidals (dSph) than in the MW if the cross section is
$p$-wave suppressed.  Measured values of $\sigma_v$ are 
less than 15 km/s in MW  dSph satellites \cite{Walker:2007ju}, whereas most estimates of
$\sigma_v$ near the GC are $\gtrsim 130\,$km/s (see for example refs.\ 
\cite{Battaglia:2005rj,Dehnen:2006cm}).  On the other hand, 
Fermi upper limits on $\langle\sigma v \rangle$ from dSph observations
are at most a factor of a few more stringent than the values of
$\langle\sigma v \rangle$ needed to fit the GCE.

\subsection{The Milky Way}
\label{sec:MW}

The Milky Way, though composed predominantly of dark matter, has inner
regions such as the bulge and bar (as well as Sagittarius A$^*$) which
are dominated by baryonic matter or otherwise do not follow an Navarro-Frenk-White (NFW) \cite{Navarro:1995iw} profile. The velocity dispersion of dark matter in the Milky Way is difficult to measure directly in the inner region, 
hence we rely on simulations and theoretical estimates.
In order to quantify the uncertainties associated with choosing a
velocity dispersion profile, we base our profiles on the results of
simulations~\cite{RomanoDiaz:2008wz} that include baryonic matter to
study the evolution of the Milky Way's profile.

If the Milky Way contained no baryonic matter, it could be suitably
modeled by an NFW profile. The resulting velocity dispersion, from
fits to the aforementioned simulation, is
\begin{align}
\sigma_v^3(r)=v_0^3\left(\frac{r}{R_s}\right)^{\chi}\left(\frac{\rho(r)}{\rho_0}\right)
\label{eq:DMB}
\end{align}
with $\chi=1.87$~\cite{Chen:2009av}. When baryons are included,
however, a slope of $\chi=1.64$ provides a better fit to the
simulations~\cite{Cirelli:2010nh}. We use the value
$v_0=130\,\rm{km}\,\rm{s}^{-1}$, consistent with the results of \cite{Battaglia:2005rj,Dehnen:2006cm}.

A second possibility that we consider is that the velocity
dispersion of the Milky Way scales as a simple power law,
\begin{align}
\sigma_{v}=v_0\left(\frac{r}{R_s}\right)^\alpha,
\label{eq:DM}
\end{align}
as suggested by the results of ref.\ \cite{RomanoDiaz:2008wz}.
A numerical fit to those results gives $\alpha\cong-1/4$~\cite{Chen:2009av} and,
using our convention, a value of $v_0=104\,\rm{km}\,\rm{s}^{-1}$,
which results in the same velocity dispersion at $r=R_\odot$ as eq.\ 
(\ref{eq:DMB}).  Ref.\ \cite{RomanoDiaz:2008wz} resolves only down 
to radii $r> 1\,$kpc, so (\ref{eq:DM}) need not hold at smaller
radii.  Nevertheless we extrapolate it to $r< 1\,$kpc to estimate
the theoretical upper bound on the predicted GCE signal, which
is greater for the ansatz (\ref{eq:DM}) than for eq.\ (\ref{eq:DMB}).
Since the observed signal is averaged over volume with $r^2$ 
weighting, the difference for the predicted GCE excess
between the two assumptions is relatively small despite the fact 
that $\sigma_v$ has very different behavior between the two as
$r\to 0$.

\subsection{Dwarf Spheroidal Galaxies}

Dwarf spheroidal galaxies tend to have relatively flat
observed  velocity
dispersion profiles out to large radii~\cite{Walker:2007ju}. We
therefore approximate them as being constant, independent of
radius.  In this case, the $J$-factor for $p$-wave annihilation
is simply proportional to that for $s$-wave. We define the
former to be
\begin{align}
J_p&\equiv\int_{\Delta\Omega}\int_{\rm l.o.s.} \rho(x)^2 
\left(\sigma_v(r)\over c\right)^2\,dl\,d\Omega'
\label{eq:J}
\end{align}
In ref.\ \cite{Ackermann:2013yva}. 
the $s$-wave $J$-factors of the 18 dwarf spheroidal galaxies for which
kinematic data was available were computed. We use these to determine
$J_p$ through the relation  
$J_p=J(\sigma_v/c)^2$. Table~\ref{tab:J} shows the velocity dispersions
and $J$-factors of the dwarf galaxies used.

\begin{table}
\begin{tabular}{|c|cccc|}
\hline
Galaxy&$\sigma_v$ (km$/$s)&~$\log_{10}J$~&~$\log_{10}J_p$~&Ref.\\
\hline
Carina&7.5&$18.1\pm0.23$&8.9&\cite{Wilkinson:2006qq}\\
Draco&13&$18.8\pm0.16$&10.1&\cite{Wilkinson:2004fz}\\
Fornax&11.1&$18.2\pm18.2$&9.3&\cite{Walker:2005nt}\\
Leo I&9.9&$17.7\pm17.7$&8.7&\cite{Koch:2006in}\\
Leo II&6.8&$17.6\pm0.18$&8.3&\cite{Coleman:2007xe}\\
Sculptor&9&$18.6\pm0.18$&9.6&\cite{Westfall:2005ji}\\
Sextans&8&$18.4\pm0.27$&9.3&\cite{Wilkinson:2006qq}\\
Ursa Minor&12&$18.8\pm0.19$&10.0&\cite{Wilkinson:2004fz}\\
Bootes I&6.6&$18.8\pm 0.22$&9.5&\cite{Belokurov:2006hf}\\
Canes Venatici I&7.6&$17.7\pm0.26$&8.5&\cite{Simon:2007dq}\\
Canes Venatcici II&4.6&$17.9\pm0.25$&8.3&\cite{Simon:2007dq}\\
Coma Berenices&4.6&$19.0\pm0.25$&9.4&\cite{Simon:2007dq}\\
Hercules&5.1&$18.1\pm0.25$&8.6&\cite{Simon:2007dq}\\
Leo IV&3.3&$17.9\pm0.28$&8.0&\cite{Simon:2007dq}\\
Segue 1&4.3&$19.5\pm0.29$&9.8&\cite{Geha:2008zr}\\
Ursa Major I&7.6&$18.3\pm0.24$&9.1&\cite{Simon:2007dq}\\
Ursa Major II&6.7&$19.3\pm0.28$&10.0&\cite{Simon:2007dq}\\
Willman 1&4.0&$19.1\pm0.31$&9.3&\cite{Willman:2010gy}\\
\hline
\end{tabular}
\caption{J-factors for dwarf spheroidal galaxies with kinematic data~\cite{Ackermann:2013yva} and velocity dispersion (with associated reference). $J$ and $J_p$ are given in $\rm{GeV}^2\,\rm{cm}^{-5}\,\rm{sr}$.}
\label{tab:J}
\end{table}

\section{Simulations and Indirect Limits}
\label{sec:simulation}

The observed gamma ray excess, if it originates from dark matter, can
be the result of annihilations to SM particles. It has been shown that
the observed flux can be fit by annihilations with a large branching
ratio to $b\bar{b}$, as would be expected for Higgs
portal dark matter \cite{Daylan:2014rsa,Calore:2014xka,Cuoco:2016jqt}. Although most
of these gamma rays are prompt (decay products of the $b$ quarks), a
significant fraction comes from inverse Compton scattering and,  to a
lesser extent, from bremsstrahlung. While the prompt signal can be
relatively easily computed, the ICS and bremsstrahlung contributions
are more involved. To this end, we use the DRAGON~\cite{Evoli:2008dv}
code to simulate cosmic ray production and propagation from dark
matter annihilations, and the GammaSky program which implements
GALPROP \cite{Strong:1998fr} to simulate the ICS and bremsstrahlung
contributions along the line of sight. GammaSky is as yet unreleased,
but some results have been given in~\cite{DiBernardo:2012zu}.

We have
modified DRAGON to account for $p$-wave annihilating dark matter,
replacing the constant cross section appearing with the DM density
by 
$\sigma v \rho(r)\rightarrow\rho(r)\,C_\sigma(\sigma_v(r)/c)^2$.
We also incorporate a generalized NFW profile 
\begin{align}
\rho(r)=\frac{\rho_0}{\left(\frac{r}{R_s}\right)^{\gamma}\left(1+\frac{r}{R_s}\right)^{3-\gamma}}
\label{eq:NFW}
\end{align} 
and the galactic diffusion parameters and magnetic field model used in
ref.\ \cite{Calore:2014xka}, corresponding to their best-fit model 
(therein called Model F).  The NFW parameters are taken to be
$\rho_0=0.3\,\rm{GeV}\,\rm{cm}^{-3}$ (giving a local DM
density of $0.4\,\rm{GeV}\,\rm{cm}^{-3}$), $R_s=20\,\rm{kpc}$, and
$\gamma=1.2$ (the best-fit value for the GCE found
in~\cite{Calore:2014xka,Daylan:2014rsa}). The electron injection
spectrum is taken from PPPC 4~\cite{Cirelli:2010xx,Ciafaloni:2010ti},
as is the  photon spectrum used in calculating the prompt
contribution.

We will focus on models in which DM
annihilates into on-shell scalar mediators $\phi$ that subsequently decay
into SM particles, primarily $b\bar{b}$. The prompt photon and electron
spectra must be boosted with respect to those from DM annihilating at
rest, to account for the velocity of $\phi$ when it decays. The decay spectrum into particles of type $i=\gamma,e$ in the
rest frame of the $\phi$ is denoted by $dN^{(\phi)}_i/dE$. It is related to the spectrum in the center of mass frame of
the $\chi\bar{\chi}$ system by \cite{Agrawal:2014oha,Cline:2015qha}
\begin{align}
\frac{dN^{(\chi)}_i}{dE}=\frac{2}{(x_+-x_-)}\int^{E\,x_+}_{E\,x_-}\frac{dE'}{E'}\frac{dN^{(\phi)}_i}{dE'},
\end{align}
where $x_\pm = m_\chi/m_\phi\pm\sqrt{(m_\chi/m_\phi)^2-1}$.
This expression assumes that the final state particles are massless, which is approximately true for the electrons as well as the photons injected from $b$ decays.

The prompt photon spectrum can be calculated independently of the
DRAGON simulation.  Its integrated spectral flux (in units of 
$\rm{photons}\,\cdot\rm{cm}^{-2}\,s^{-1}$) is given by
\begin{align}
\frac{d\Phi_{\rm prompt}}{dE}=\frac{C_\sigma}{8\pi m_\chi^2}\frac{dN_\gamma}{dE}\times J_p,
\label{eq:prompt}
\end{align}
with $J_p$ defined in eq.\ (\ref{eq:J}). The total observed spectrum 
is equal to the sum of ${d\Phi_{\rm prompt}/dE}$
 and the ICS+Bremsstrahlung spectrum determined from the simulations.

\subsection{Simulation Results}

\begin{figure*}
\begin{center}
\includegraphics[scale=0.4]{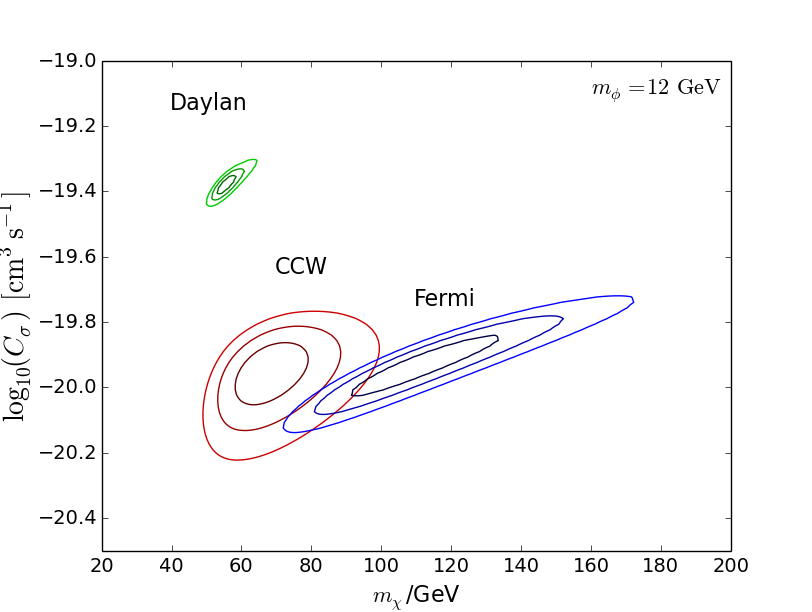}
\includegraphics[scale=0.4]{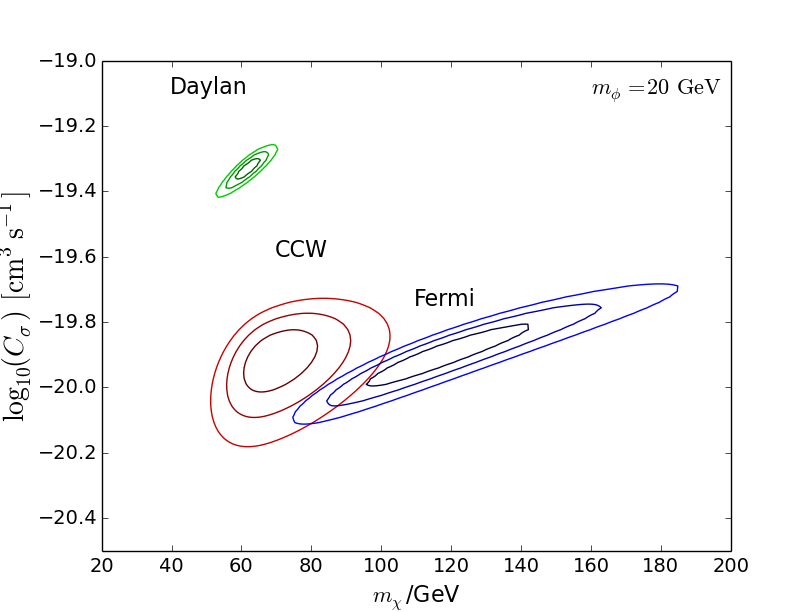}
\includegraphics[scale=0.4]{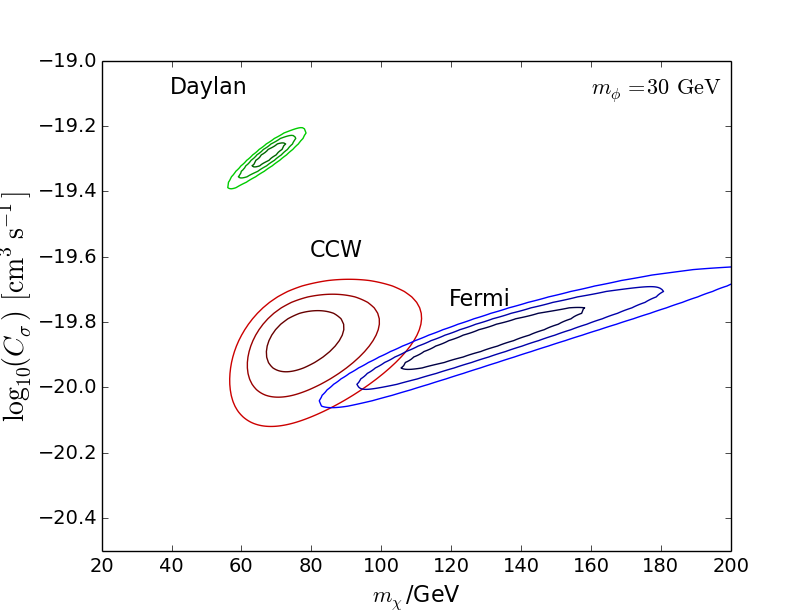}
\includegraphics[scale=0.4]{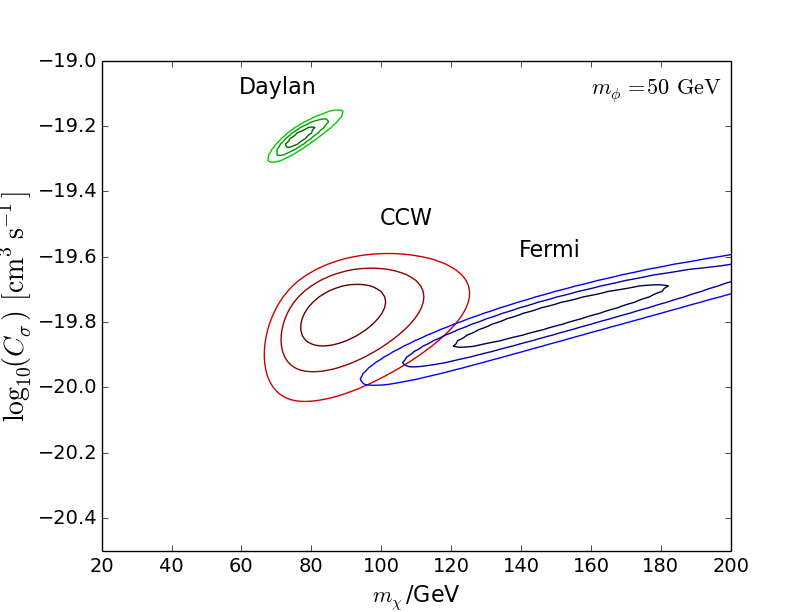}
\caption{1$\sigma$, 2$\sigma$, and 3$\sigma$ contours for the CCW~\cite{Calore:2014xka}, Daylan 
{\it et al.} \cite{Daylan:2014rsa}, 
and Fermi Collaboration~\cite{TheFermi-LAT:2015kwa} data. 
The results are shown for annihilation into on-shell scalar 
mediators, followed by decay into $b\bar{b}$, with a mediator 
mass of $m_\phi=12$, $20$, $30$, and $50\,\rm{GeV}$.  Shaded regions
in upper left corner indicate the constraint from the Virgo cluster.}
\label{Fig:contours_noDwarfs}
\end{center}
\end{figure*}

\begin{figure}
\begin{center}
\includegraphics[scale=0.45]{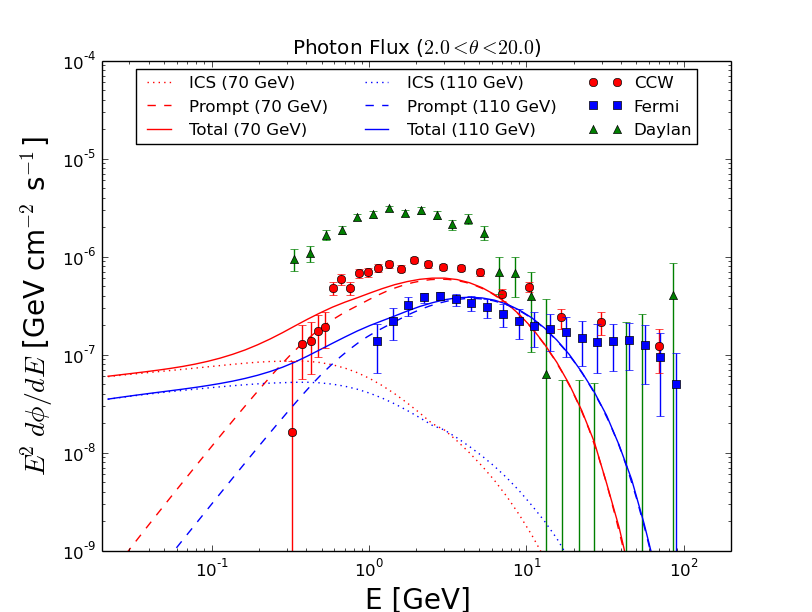}
\caption{Simulated observed photon energy flux for $p$-wave
annihilating dark matter with
$m_\chi=70\,\rm{GeV}$ (red, upper curved) or $m_\chi=110 \,\rm{GeV}$
(blue, lower curves), mediator mass
 $m_\phi=12\,\rm{GeV}$ and cross section coefficient $C_\sigma=10^{-20}\,\rm{cm}^3s^{-1}$. The observed region is the disk-like region $2.0<\theta<20.0$, where $\theta$ is the viewing angle as measured from the galactic center.
The ICS+bremsstrahlung (dotted) and prompt (dashed) components are shown
individually. Also shown are the three datasets of
observed fluxes; the values of $m_\chi$ are chosen to demonstrate
the best fits to two of the individual datasets.}
\label{Fig:Flux}
\end{center}
\end{figure}

We simulated the gamma ray flux for a range of dark matter masses
($20\,\rm{GeV}\leq m_\chi\leq 200\,\rm{GeV}$) and compared the results
to the GCE signals estimated in refs.\
\cite{Calore:2014xka,Daylan:2014rsa,FermiData}. The best fit regions
are presented in fig.\ \ref{Fig:contours_noDwarfs}, which show the
confidence intervals in the $C_\sigma$-$m_\chi$ plane for
four different values for the
mediator mass, $m_\phi = 12,\,20,\,30,\,50\,$GeV. The contours are generated by minimizing the $\chi^2$ of our simulated spectrum with respect to each dataset in the $C_\sigma$-$m_\chi$ plane, and contours are then drawn at $\chi^2_{\rm min}+2.30$, $\chi^2_{\rm min}+6.18$, and $\chi^2_{\rm min}+11.83$, corresponding to $1\sigma$, $2\sigma$, and $3\sigma$. The minimum values
of $\chi^2$ and the corresponding model parameters are given in
table \ref{tab:chi2}, which shows that the fit results are relatively
insensitive to the mediator mass (the fits to the Fermi data display a mild
preference for heavier mediators).  Reasonably good fits to the Fermi
and CCW data sets are obtained, with
\begin{align} 
   m_\chi \sim 90\,\rm{GeV},\quad
   C_\sigma \sim 10^{-20}\,\rm{cm}^3\,\rm{s}^{-1}
\label{eq:fit}
\end{align}
whereas the fit to the Daylan {\it et
al.} data is poor.   The data are compared to the
simulated observed spectrum from the GC in
fig.\ \ref{Fig:Flux} for representative values of $m_\chi$ and
$C_\sigma$, taking a mediator mass of
$m_\phi=12\,\rm{GeV}$.

\begin{figure}[t]
\begin{center}
\includegraphics[scale=0.4]{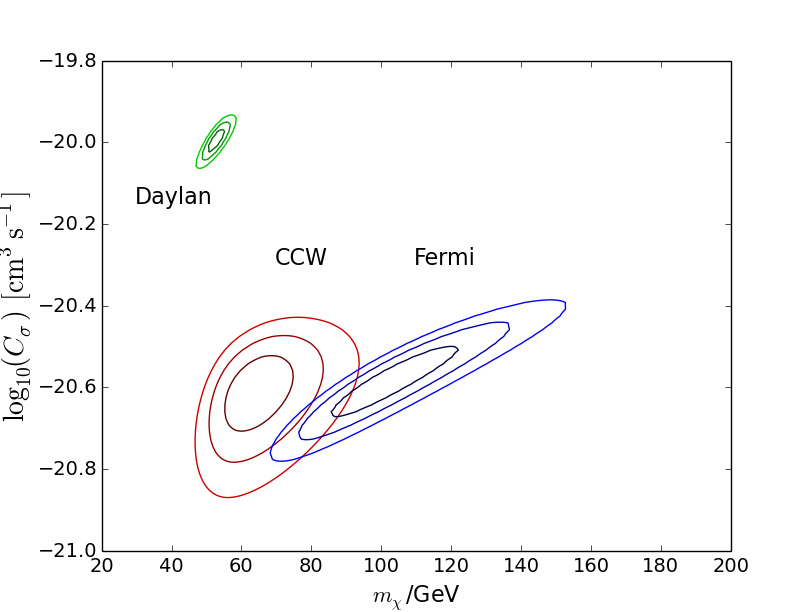}
\caption{Like fig.\ \ref{Fig:contours_noDwarfs}, but using the 
velocity dispersion profile in equation~\ref{eq:DM}, with a mediator mass of $m_\phi=12$.}
\label{Fig:contours_otherVels}
\end{center}
\end{figure}

The previous results are based upon the
assumption of eq.\ (\ref{eq:DMB}) for the DM velocity disperion in the
MW.  The effect of using higher $\sigma_v$, using eq.\
\ref{eq:DM}, is shown in fig.\ \ref{Fig:contours_otherVels}, which
results in somewhat lower central values of 
$C_\sigma\sim 0.2\times 10^{-20}\,\rm{cm}^3\,\rm{s}^{-1}$ for the 
cross section  and $m_\chi\sim 80\,$GeV for the mass.

\begin{table}
\begin{tabular}{|c|ccc|ccc|ccc|}
\hline
&\multicolumn{3}{c|}{CCW ($N=21$)}&
\multicolumn{3}{c|}{Fermi ($N=20$)}&\multicolumn{3}{c|}{Daylan ($N=25$)}\\
$m_\phi$&$\chi^2_{\rm min}$&$m_\chi$&$\log_{10}C_\sigma$&$\chi^2_{\rm min}$&$m_\chi$&$\log_{10}C_\sigma$&$\chi^2_{\rm min}$&$m_\chi$&$\log_{10}C_\sigma$\\
\hline
12&29.7&68&-20.0&24.9&109&-19.9&54.1&56&-19.4\\
20&29.9&70&-19.9&23.7&116&-19.9&65.3&62&-19.3\\
30&29.9&76&-19.9&22.7&128&-19.9&71.3&67&-19.3\\
50&30.6&88&-19.8&22.0&146&-19.8&76.8&76&-19.2\\
\hline
\end{tabular}
\caption{Minimum $\chi^2$ values for fits to the three datasets,
(number of data points $N$ indicated). 
Masses are in GeV and $C_\sigma$ is in $\rm{cm}^3\rm{s}^{-1}$. 
The confidence regions are shown in fig.\ 
\ref{Fig:contours_noDwarfs}.}
\label{tab:chi2}
\end{table}

\begin{figure}[t]
\begin{center}
\includegraphics[scale=0.4]{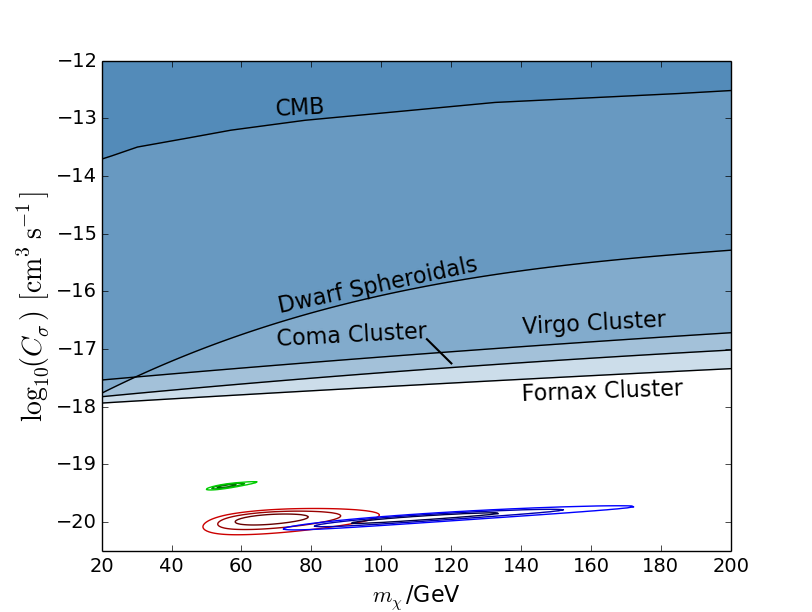}
\caption{Like figure~\ref{Fig:contours_noDwarfs}, including $95\%$
C.L. upper
limits on $C_\sigma$ from the five most constraining dwarf spheroidals, the Virgo, Fornax, and Coma clusters, and the CMB. The fits to
the GCE for $p$-wave annihilating dark matter are well below the
limits.  The CMB constraint is
taken from ref.\ \cite{Diamanti:2013bia}, for the case of
annihilations to $e^+e^-$.}
\label{Fig:contours_segue}
\end{center}
\end{figure}

\subsection{Limits from Dwarf Spheroidal Galaxies}
\label{sec:dwarflimit}

An upper limit on the gamma-ray flux from DM annihilation in 18 dwarf
spheroidal galaxies with kinematic data has been determined using
Fermi-LAT data~\cite{Ackermann:2013yva}. This can be used in
conjunction with the $J$-factors presented in table \ref{tab:J} to obtain
an upper limit on $C_\sigma$. The strongest such constraint comes from
the dwarf galaxy Draco. At a distance of 80 kpc and with a relatively
large $J$-factor and high velocity dispersion, it would
be the most likely to exhibit signs of $p$-wave annihilating dark
matter. 

Ackermann \emph{et al.} give the combined limit on $\langle\sigma
v\rangle_{b\bar{b}}$ (annihilation into $b\bar b$) at 95\% C.L. for 15
dwarf spheroidal galaxies. In our model, DM annihilates to
$bb\bar b\bar b$, leading to a different gamma-ray spectrum, but  in this
section and the next we assume the resulting limit on the annihilation
rate in both cases is approximately the same.  (Note that the total
energy deposition in the two cases is the same.)  

The previously derived limit assumes $s$-wave annihilation and 
therefore cannot be directly converted into a limit from $p$-wave
annihilation, as the different velocity dispersions of the dwarf
spheroidals would have to be taken into account individually. If,
however, we make the simplifying assumption that all dwarf spheroidal
galaxies have velocity dispersions equal to the greatest value (that
of Draco, with $\sigma_v=13$ km/s), we can then use
equation~\ref{eq:Csigma} to directly convert the limit to one on
$C_\sigma$. This will lead to a constraint that is slightly more 
stringent than the true value, but sufficient for our purpose of
showing that there is no tension with the GCE.
The resulting upper limit on $C_\sigma$ as a function of $m_\chi$
is shown in figure~\ref{Fig:contours_segue}, 
along with the GCE best-fit
regions.   The weaker CMB constraint from energy injection at
recombination~\cite{Diamanti:2013bia} (also discussed in section \ref{cmb-constraints}) is also indicated there.  

We see that the assumption of $p$-wave annihilation rather than
$s$-wave completely eliminates the tension between the dwarf
spheroidal constraints and the GCE. The former are softened by a
factor of $\sim\sigma^2_{v,\rm{dwarf}}/\sigma^2_{v,\rm{MW}}\sim
(13/130)^2\sim 10^{-2}$ relative to the GCE signal. 
The constraints depend on the
velocity dispersion profile assumed for the dwarfs, but even taking
into account the uncertainties, the
limiting cross section from dwarf spheroidal galaxies is far above
the values required to explain the GCE.

\subsection{Galaxy Cluster Limits}
Searches for gamma rays from galaxy clusters can place more
stringent constraints on our scenario. Although dwarf spheroidal 
constraints were weakened due to their smaller velocity dispersion,
the converse is true for clusters: their larger velocity dispersions 
amplify the signal from $p$-wave annihilating dark matter, relative to
smaller systems.

Observations of the Coma
\cite{Rephaeli:2015nca} and Virgo \cite{Ackermann:2015fdi} clusters
have recently been analyzed by the Fermi-LAT Collaboration.  The first of
these references gave no limits on annihilating dark matter, while the
second did so for $s$-wave annihilations. We therefore derive the bound on $p$-wave annihilating DM arising from  the latter.  For
this purpose we adopt a value for the velocity dispersion of $643$ km/s for
Virgo~\cite{Girardi:1995iy}.

Limits on $\langle\sigma v \rangle_{b\bar{b}}$ are derived at 95\% C.L. for
the Virgo cluster in~\cite{Ackermann:2015fdi}, using a background model
taking into account all Fermi 2-year catalog point sources as well as
diffuse galactic and extragalactic spectra. We have converted them directly
into limits on $C_\sigma$ using equation~\ref{eq:Csigma}, with one caveat:
dark matter substructure---subhalos residing within the larger host halo---is
expected to significantly boost the signal strength from $s$-wave dark
matter annihilation over what would be expected from the host halo
alone. The constraints in ref.\ \cite{Ackermann:2015fdi} for the more conservative limit given
assume a boost factor of $b=33$ from the substructure of the
cluster. The substructure is not expected to have the same velocity
dispersion as the host halo however, making the simple rescaling described in the previous section
inapplicable for $p$-wave annihilation. 
Subhalos generically have a significantly smaller velocity dispersion
than the host halo, due to the fact that the velocity dispersion depends on the total mass of the subhalo where the dark matter is virially bound, not on that of the host halo. This can be seen in simulations such as RHAPSODY~\cite{Wu:2012wu,Wu:2012dg}, in which the number of galaxy cluster subhalos is found to drop off sharply with increasing maximum circular velocity (a power law index of -2.95) with no subhalos exceeding a third of the host halo's maximum circular velocity. The contribution to the signal from subhalos is therefore weakened due to the velocity dependence of the annihilation cross section, offsetting the gains that come from the increased dark matter density. Ultimately, we choose a conservative approach and rescale the limits from~\cite{Ackermann:2015fdi}
by a factor of $b+1$ to remove the boost from the substructure for a self-consistent limit. The upper limits on $C_\sigma$ are
shown in fig.~\ref{Fig:contours_segue}.

Similar limits have been found for several other clusters, including
Coma and Fornax, using earlier Fermi data~\cite{Ackermann:2010rg}. The Fornax cluster was subsequently reanalyzed with
specific attention to the effects of subhalos and contraction due to
baryonic infall~\cite{Ando:2012vu}, leading to a more stringent upper
bound on $\langle\sigma v\rangle_{b\bar{b}}$. As with the Virgo
cluster, from this work we use the conservative limits neglecting the effect
of substructure, which in ref.\ \cite{Ando:2012vu}
are given alongside the more
optimistic limits. We convert the constraints on
Coma~\cite{Ackermann:2010rg} (which does not account for substructure)
and Fornax~\cite{Ando:2012vu} directly into limits on
$C_\sigma$, using velocity dispersions of $913$
km/s~\cite{Girardi:1995iy} and $370$ km/s~\cite{Drinkwater:2000dr}
respectively; these are also included in
fig.~\ref{Fig:contours_segue}.

Although our best-fit parameters are consistent with older bounds from
the Virgo and Fornax clusters, more recent observations of the
Coma cluster are expected to give more stringent constraints due to
its high dark matter density and larger velocity dispersion. Currently
there are no  limits on dark matter annihilation rates from the more
recent observations, and such a study is beyond the scope of the
present work.

\subsection{Isotropic Gamma-Ray Background}

The isotropic gamma-ray background (IGRB) could further constrain our
scenario. As part of the DM annihilation contribution to this signal could be from even larger halos than the ones surrounding the galaxy clusters we considered in the previous section, it is possible that it could be enhanced if the annihilation cross section is velocity dependent. The most recent measurements of the IGRB can constrain the
$s$-wave annihilation cross-section to $\langle\sigma v\rangle\lesssim
10^{-24}\,\rm{cm}^3\,\rm{s}^{-1}$ for conservative limits and
$\langle\sigma v\rangle\lesssim \langle\sigma v \rangle_{\rm thermal}$
for more optimistic limits corresponding to our adopted best-fit
value of $m_\chi=80\,\rm{GeV}$~\cite{Liu:2016ngs}.
Converting these
limits into constraints on $C_\sigma$ is not a simple matter, as arriving at an expected IGRB signal requires taking into account how the velocity dispersion varies for halos of different sizes and at different redshifts.
Such a detailed analysis is beyond the scope of this work but 
would be interesting for future investigation.

\section{Relic abundance from decaying dark matter}
\label{sec:relic}

An immediate problem with $p$-wave annihilating DM in the galactic
center is that the corresponding cross section in the early universe
would have been orders of magnitude greater, due to the larger
relative velocities, leading to a highly suppressed relic density.
The form of the Boltzmann equation which describes the time evolution of the number density
for Dirac dark matter  $\chi$ is
\be
\frac{d n_\chi}{d t} + 3 H n_\chi = -\langle 
\sigma v \rangle \left(n_\chi n_{\bar{\chi}} - n_\chi^{\rm EQ} 
n_{\bar{\chi}}^{\rm EQ} \right) \, ,
\ee
where $n^{\rm EQ}$ is the number density of a particle in thermal equilibrium with the photon bath.
The equation for the evolution of the number density of the antiparticle $\bar{\chi}$ is of the same form. We assume
that there is no asymmetry between $n_\chi$ and $n_{\bar{\chi}}$, and therefore the total number
density $n = n_\chi + n_{\bar{\chi}}$ is given by
\be
\label{Boltz}
\frac{d n}{d t} + 3 H n = - \frac {\langle \sigma v \rangle}{2} 
\left(n^2 - n_{\rm EQ}^2 \right) \,. 
\ee
Following the procedure of ref.\ \cite{Kolb:1990vq}, 
an approximate solution of the Boltzmann equation for the relic
density is given by 
\be
\label{eq:omegah2}
\Omega_\chi h^2 = \frac{\rho_\chi}{\rho_c} h^2 = 2.14 \times 10^9 
\frac{(n+1) x_f^{(n+1)}}{\left(g_{\star S} / g_{\star}^{1/2} \right) 
M_{\rm Pl}\, \sigma_0} \, {\rm GeV}^{-1} \, .
\ee
where $x_f = m_\chi / T_f$, $T_f$ is the freeze-out temperature, 
and the effective degrees of freedom $g_{\star}$ and $g_{\star S}$ 
are evaluated at $T_f$. The thermally averaged cross section takes 
the form $\langle \sigma v \rangle = \sigma_0\, x_f^{-n}$; hence
$n=1$ and $\sigma_0 = 3\,C_\sigma$ for our $p$-wave annihilation 
scenario where
\be
\langle \sigma v \rangle = 3\, C_\sigma \frac{T}{m_\chi}
\ee
An approximate solution for $x_f$ is given by 
\bea
x_f &=& \ln y_f
-(n+1/2)\ln\ln y_f,\\
y_f &=& 0.038\,(n+1)g_*^{-1/2} M_{\rm Pl}\, m_\chi\, \sigma_0\nonumber
\label{eq:xf}
\eea
Our fiducial fit, eq.\ (\ref{eq:fit}), implies
\be
\label{xfeq}
x_f=32.3\quad
\Omega_\chi= {\rho_\chi\over \rho_c} = 3.6\times10^{-5}
\ee
to be compared to the observed value $\Omega_{DM} = 0.26$ \cite{Ade:2015xua}.
Hence the 
thermally produced abundance  is approximately 7000 times too small;
we need a nonthermal production mechanism. 

\subsection{Decaying dark matter}

A conceptually simple solution, similar to the superWIMP model proposed in \cite{Feng:2003xh}, is to suppose that today's dark matter
$\chi$ is the product of a heavier metastable 
state $\psi$, that decayed into $\chi$ at temperatures
below freeze-out of $\bar{\chi}\chi$ annihilations.
For $m_\chi\sim 90{\rm\,GeV}$, this occurs at $T_f\sim m_\chi/32\sim
3{\rm\,GeV}$ according to (\ref{xfeq}).  Hence we need for $\psi$ to
have a lifetime exceeding $10^{-6}\,$s.  
Such long lifetimes are suggestive of an analog of weak
interactions in the dark sector.  We consider representative effective
interactions giving rise to decays $\psi\to\chi\gamma$,
$\psi\to\chi e^+ e^-$ or $\psi\to\chi b\bar b$, of the form
\be
\label{eq:eff_op}
{1\over\Lambda_\gamma}\,\bar\chi \sigma_{\mu\nu}\psi\,F^{\mu\nu},\quad
{(\bar\chi \gamma^\mu\psi)(\bar e\gamma_\mu e)\over\Lambda_e^2},\quad
{(\bar\chi \gamma^\mu\psi)(\bar b\gamma_\mu b)\over\Lambda_b^2},
\ee
where $\Lambda_{e,\gamma,b}$ are heavy scales. Each operator is also accompanied by its Hermitian conjugate, which leads to decays of $\bar\psi$.  These decay channels are
chosen to illustrate constraints that can arise from big bang
nucleosynthesis (BBN) and the cosmic microwave background (CMB).  
An alternate channel $\psi\to\chi \nu\bar\nu$ would be safe from these
constraints.
The decay rates corresponding to the first two operators are given by
\be
\label{rates}
	\Gamma_\gamma = {4\,\delta m^3\over \pi\Lambda_\gamma^2},\quad
	\Gamma_{ee} \cong {\delta m^5\over 60\pi^3 \Lambda_e^4}
\ee
where the mass splitting $\delta m = m_\psi-m_\chi$ is considered to 
be much less than $m_\psi\cong m_\chi$, but greater than $2 m_e$
for decays into electrons.  (We ignore phase space effects in the
small region of parameter space where $\delta m\gtrsim 2 m_e$.) 
For the third operator, we are interested in larger mass splittings
since $\delta m$ must be at least $2 m_b$.  We use numerical 
results for its decay rate where needed.  A fairly good fit is given
by $\Gamma_b \cong A_0( m_\psi^{A_1}-m_\chi^{A_1})^{A_2}/\Lambda_b^4$
where for $\Gamma_b,\,m_{\psi,\chi}$ in GeV units, $A_{0,1,2} =
(3.60,\,1.33,\,2.30)$.

To obtain the relic density of the parent particle $\psi$, we
assume for definiteness an effective interaction
\be
\label{ann_op}
\quad
{(\bar\psi \gamma^\mu\psi)(\bar f\gamma_\mu f)\over\Lambda_f^2},
\ee
giving rise to $\psi\bar\psi\to f\bar f$, where $f$ can be a 
light fermion of the standard model or in a hidden sector.  The annihilation 
cross section for $\psi\bar\psi\to f\bar f$ is 
\be
\label{eq:chi_ann}
	\langle\sigma v\rangle \cong {m_\psi^2\over
	\pi\Lambda_f^4}
\ee

\begin{figure*}[t]
\begin{center}
\includegraphics[scale=0.7]{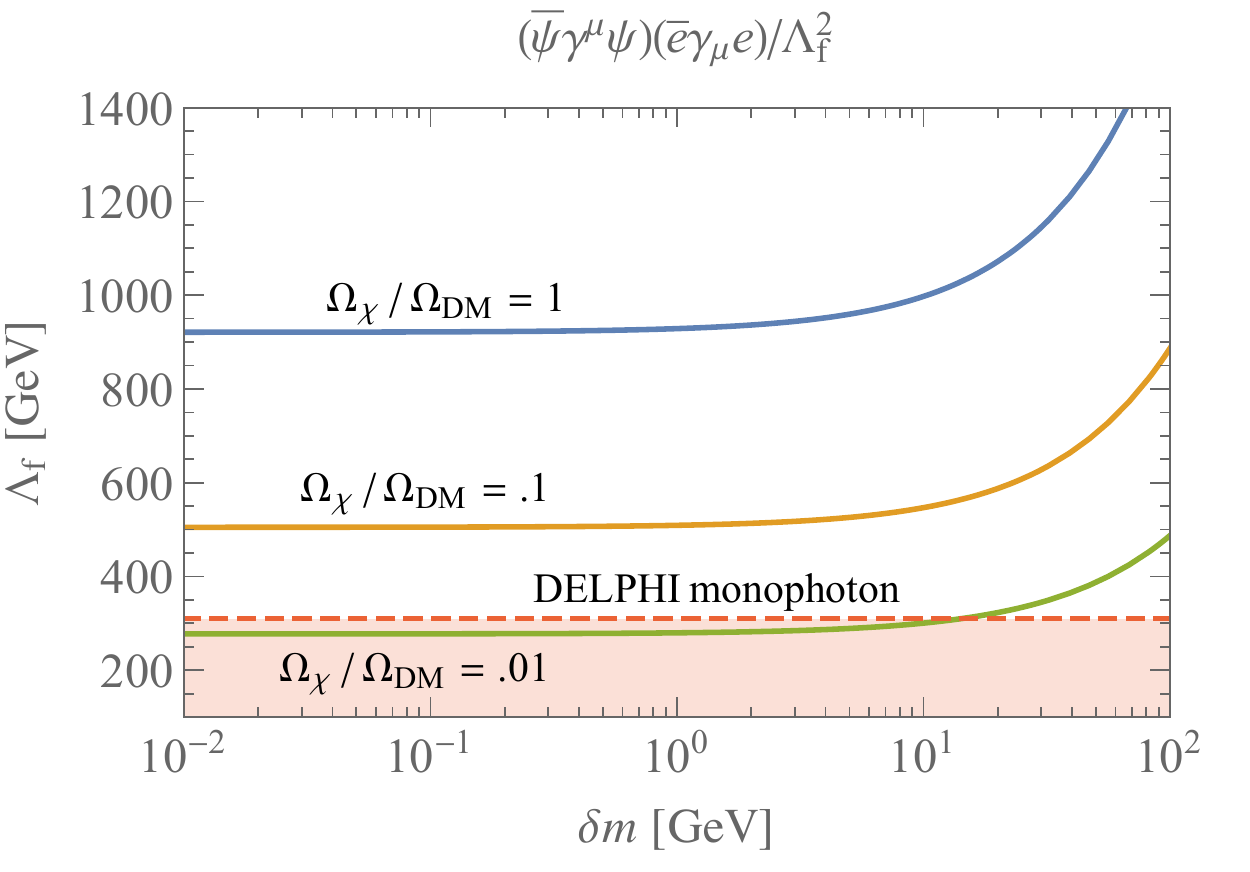}
\includegraphics[scale=0.7]{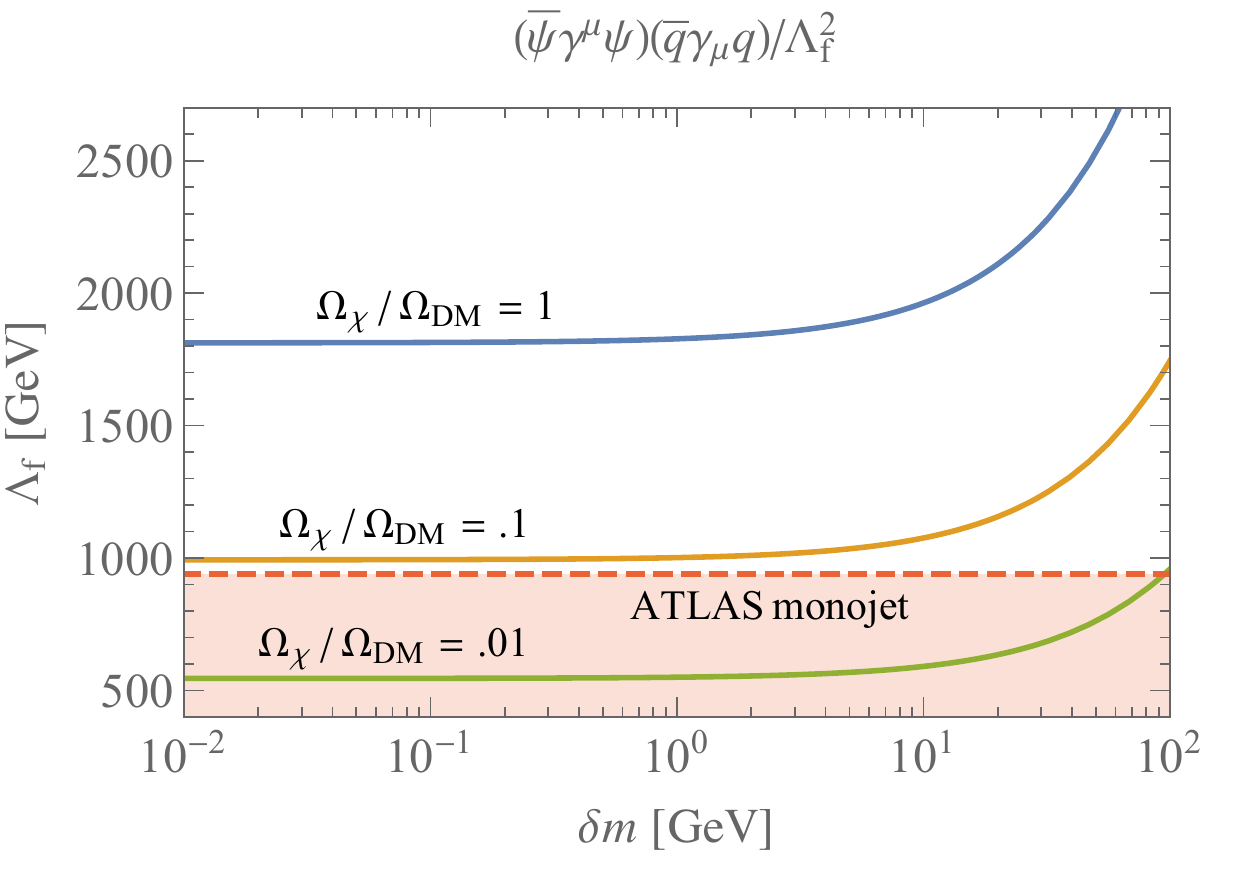}
\caption{\label{fig:relic} Contours of constant relic density for a
dark matter mass of $m_\chi = 90$~GeV assuming that the $\psi \chi$ coannihilation rate is negligible relative to the $\psi \bar \psi$ annihilation rate. In the left plot, $\psi$ and $\bar \psi$ couple to $e^{+} e^{-}$ and in the right they couple to quarks. The shaded region in the left plot is excluded at 90\% C.L. from a DELPHI search for monophotons, while an ATLAS search for monojets excludes the shaded region in the right plot at 95\% C.L.}
\end{center}
\end{figure*}

To determine the relic density in this scenario, we again use
eqs.\ \ref{eq:omegah2} and \ref{eq:xf} but now with $n = 0$, since
the vector current operators of eq.\ (\ref{eq:eff_op}) lead to $s$-wave
annihilation. The ultimate 
relic density of $\chi$ particles is related to the prior 
abundance of $\psi$ by $\Omega_\chi
= (m_\chi / m_\psi)\, \Omega_\psi$. Curves of constant $\Omega_\chi$
in the $\delta m$-$\Lambda_f$ plane for $m_\chi = 90$ GeV are shown in figure
\ref{fig:relic}. Here we consider two different scenarios: $\psi \bar \psi$ annihilations
to electrons and positrons and to quark-antiquark pairs.
Large mass splittings $\delta m \gtrsim 1\,$GeV
lead to a reduction in $\Omega_\chi$ that must be compensated by
reducing the cross section by increasing $\Lambda_f$.
These estimates assume that coannihilations $\psi\chi\to \bar f f$ as well as inelastic scatters $\psi f \to \chi f$ are
unimportant for determining the DM relic density.  This will be true
(as we explore in detail in the following subsections)
as long as $\Lambda_e\gg \Lambda_f$, which is also consistent with
the need for $\psi$ to be relatively long-lived.
For small $\delta m \lesssim 1\,$GeV,  the desired relic density 
for $\psi$ and $\chi$ is independent of $\delta m$ and requires $\Lambda_f\cong
920\,$GeV when $\psi$ and $\bar \psi$ couple to $e^{+} e^{-}$ and $\Lambda_f\cong
1810\,$GeV when they couple to $q \bar q$. 

\subsection{Coannihilations}

Coannihilation processes can reduce
the relic density of $\psi$, which was assumed to be a small 
effect in the 
previous treatment. When
the splitting between $m_\psi$ and $m_\chi$ is small, \
leading to $n_\psi \approx n_\chi$, the effect can be estimated by
replacing $\langle \sigma v \rangle$ in eq.~\ref{Boltz} with \cite{Griest:1990kh}
\be
\langle \sigma_{\rm eff}\, v \rangle = 
\langle \sigma_{\psi \bar\psi \rightarrow X\bar X}\, v 
\rangle + \langle \sigma_{\psi \chi \rightarrow X\bar X}\, v \rangle \, .
\ee
Here X represents any standard model particle, so the first term in
the above equation is the total $\psi \bar \psi$ annihilation cross
section and the second is the total $\psi \chi$ coannihilation cross
section. Eq.~(\ref{Boltz}) with this effective cross section only 
describes the number density of $\psi$ until the freeze-out of this 
species, since after that point decays and inelastic scatterings can have a significant impact on the $\psi$ number density. In this low mass splitting limit, the relevant $\psi \chi \to
f \bar f$ coannihilation cross section for the 4-fermion operator in
eq.\ (\ref{eq:eff_op}) has the same form as eq.\ (\ref{eq:chi_ann}), 
while the
dipole operator gives \cite{Weiner:2012cb}
\be
\langle \sigma_{\chi \psi \to f \bar f}\, v \rangle = \frac{4 \alpha Q_f^2} {\Lambda_\gamma^2} \, ,
\ee 
where $Q_f$ is the electric charge of the fermions in the final state
and $\alpha$ is the fine-structure constant.

In the scenario where the relic density is determined entirely by
coannihilation processes, \textit{i.e.,} when the operator in eq.\
(\ref{ann_op}) is not present, the correct relic density requires
$\Lambda_e \ge 920$ GeV or $\Lambda_\gamma \ge 8000$ GeV. These are
lower bounds, since increasing the strength of coannihilation
processes would lead to underproduction of DM, while the larger relic
density induced by decreased coannihilation can be
offset by increased $\psi \bar\psi$ annihilation. 

\begin{figure*}
\begin{center}
\centerline{
\includegraphics[scale=.43]{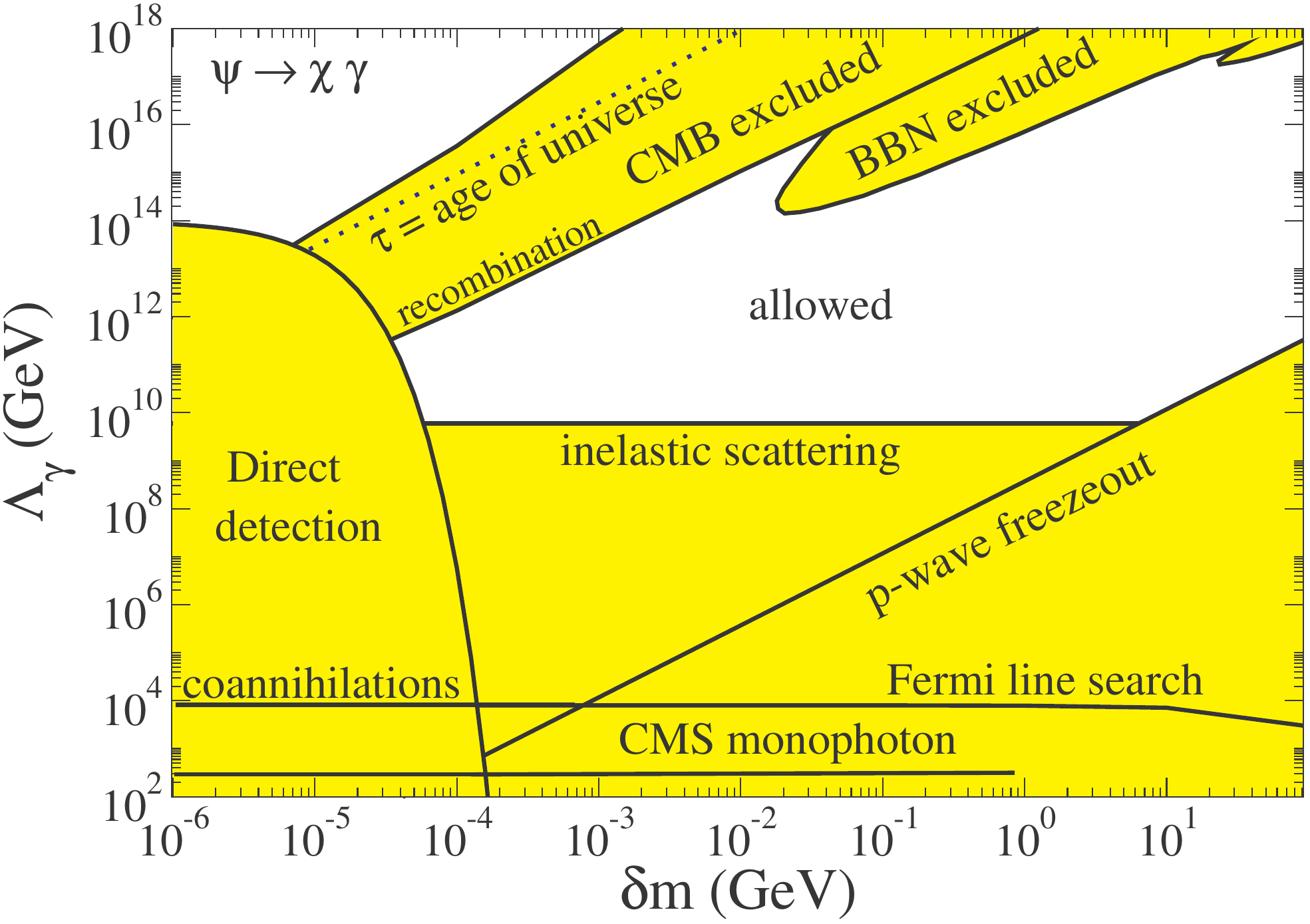}
\hfill
\includegraphics[scale=.43]{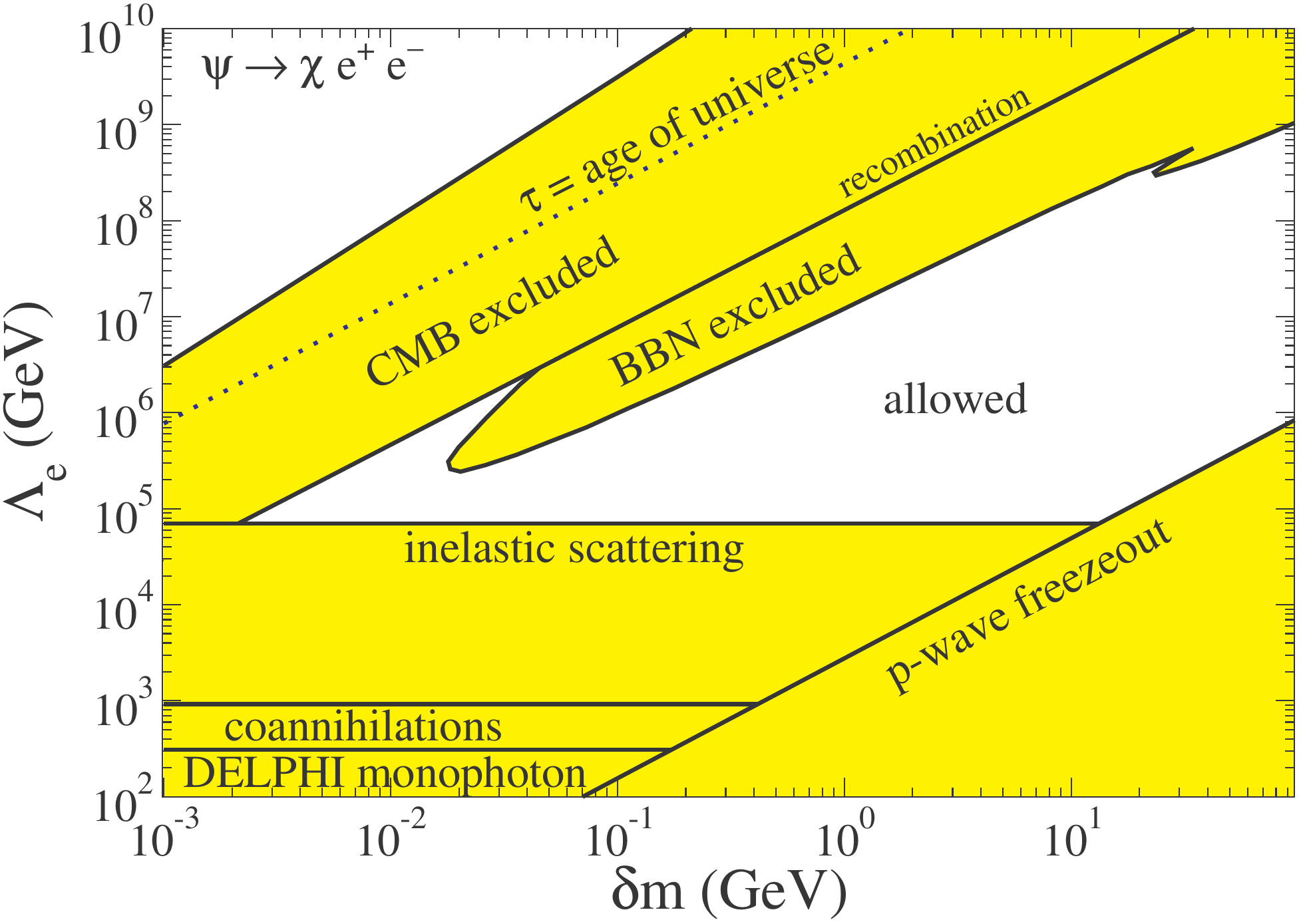}}
\caption{\label{Fig:Lifetime} 
Left: Excluded (shaded) and allowed (unshaded) regions of parameter space for $\psi\to\chi\gamma$
decays in the 
$\delta m$-$\Lambda_\gamma$ plane. In the upper-left regions, the 
lifetime of 
the $\psi$ is too great, causing its decays to interfere with BBN,
CMB, or exceed the age of the universe; in the lower right regions 
$\psi$ decays before the $\chi\bar\chi$ annihilations freeze out, 
erasing any excess above the standard $\chi$ relic abundance 
produced via pair annihilation.
Right: corresponding result for $\psi\to\chi e^+e^-$.  Dark matter
mass $m_\chi = 90\,$GeV was assumed for determining the number
density of decaying $\psi$ particles. }
\end{center}
\end{figure*}

The resulting limits are
shown in fig.~\ref{Fig:Lifetime}. Decays of $\psi$ to $b$-quarks
require a relatively large mass splitting and consequently a
more sophisticated calculation than the one described here, but the
limits on $\Lambda_b$ from suppressing coannihilations are
greatly subdominant to those from demanding that $\psi$ decays 
after $\chi$ freeze-out. We also note that the operators in
eq.\ (\ref{eq:eff_op}) lead to additional annihilation processes from the
ones we have considered above, including $\psi \bar \psi \to \bar f f
\bar f f$ for the four-fermion operator as well as $\psi \bar \psi \to
\gamma \gamma$ and $\psi \bar \psi \to \gamma \phi$ for the magnetic
dipole operator. We have checked that these are negligible when
the other constraints considered are satisfied.

\subsection{Inelastic scattering}

A further requirement for consistency of our relic density
determination is that inelastic scatterings $\psi f\to\chi f$ induced
by the decay operators (\ref{eq:eff_op}) are not
important during the epoch between $\psi$ freeze-out and the significantly later $\chi$
freeze-out. Otherwise further depletion of the
final abundance would occur due to scattering-induced $\psi\to\chi$
transitions followed by $\chi\bar\chi$
annihilations.  This leads to the criterion
\be
\label{inelastic_bound}
	(n_e+n_{\bar e})\langle \sigma v\rangle_{\psi e\to \chi e}
	< n_\psi \langle\sigma v\rangle_{\psi\bar\psi\to f\bar f}
\ee
for the $(\bar\chi\gamma^\mu\psi)(\bar e\gamma_\mu e)$ operator. We
ignore the effect of $\chi \to \psi$ transitions because the number
density of $\chi$ relative to $\psi$ is extremely suppressed in our
scenario at $\chi$ freeze-out.
For $m_\chi=90\,$GeV, freeze-out occurs at $T_f\cong
3\,$GeV, for which it is sufficient to compute the inelastic cross
section in the elastic limit $\delta m = 0$, and also 
approximating $m_e\cong 0$.  We find that
\be
	(\sigma v)_{\psi e\to \chi e} \cong {E_e^2\over
	2\pi\Lambda_e^4}
\ee
at the relevant energies.  Performing the thermal average over
electron energies gives $\langle E_e^2\rangle\cong 12.9\,T_f^2$,
and we find from (\ref{inelastic_bound}) the limit
\be
	\Lambda_e \gtrsim 1.9\, x_f^{-1/2}\, Y_{\psi/e}^{-1/4}\Lambda_f
	\cong 70\,{\rm TeV}
\label{Lelimit}
\ee
where $Y_{\psi/e}\cong 3.4\times 10^{-11}$ is the abundance of 90\,GeV
DM relative to electrons at $T_f$. 

From the magnetic dipole operator, one has photon-mediated
scattering from all charged particles that are in equilibrium at
$T_f\sim 3\,$GeV, which we take to be $f=e,\mu,\tau,u,d,s,c$ plus
their antiparticles.  The
cross section has a logarithmic infrared divergence in the limit 
$m_f\,\delta m\to
0$ from low-angle scattering.  For $m_f=m_e$, 
it is regulated more effectively by Debye screening than by the
small value of $m_e\, \delta m$, giving
\bea
	(\sigma v)_{\psi f\to \chi f} &\cong& {Q_f^2\, e^2\over
\pi\Lambda_\gamma^2}\left(2\left({m_\chi^2\over s}-1\right) \right.\nonumber\\
	&+&\left. \ln\left(
	{1+ {(s - 2 m_\chi^2)^2\over s\, m_D^2}}\right)
	\right)
\label{svmag}
\eea	
For simplicity we cut off the
divergence for all species using the Debye mass 
$m_D=(\sum_f Q_f^2\, n_f/T)^{1/2}\cong 1.5\, e\,T
\cong 1.4\,$GeV. The thermal average of (\ref{svmag}) is
$0.13/\Lambda_\gamma^2$ for the parameters of interest.  The resulting 
bound analogous to (\ref{Lelimit}) is
\be
	\Lambda_\gamma \gtrsim 4\times 10^9\,{\rm GeV}
\ee
No similar constraint arises for $\Lambda_b$ since $b$ quarks are not
present in the plasma at temperature $T_f$.

The bounds on $\Lambda_e$ and $\Lambda_\gamma$ are shown in
fig.~\ref{Fig:Lifetime}. In both cases the limits derived from 
suppressing inelastic scattering are much stronger than those
from suppressing coannihilation processes. This is because the
number density of relativistic standard model scattering partners is
much greater than the Boltzmann-suppressed number density of $\chi$ at
$\psi$ freeze-out.

\begin{figure}
\begin{center}
\centerline{
\includegraphics[scale=.43]{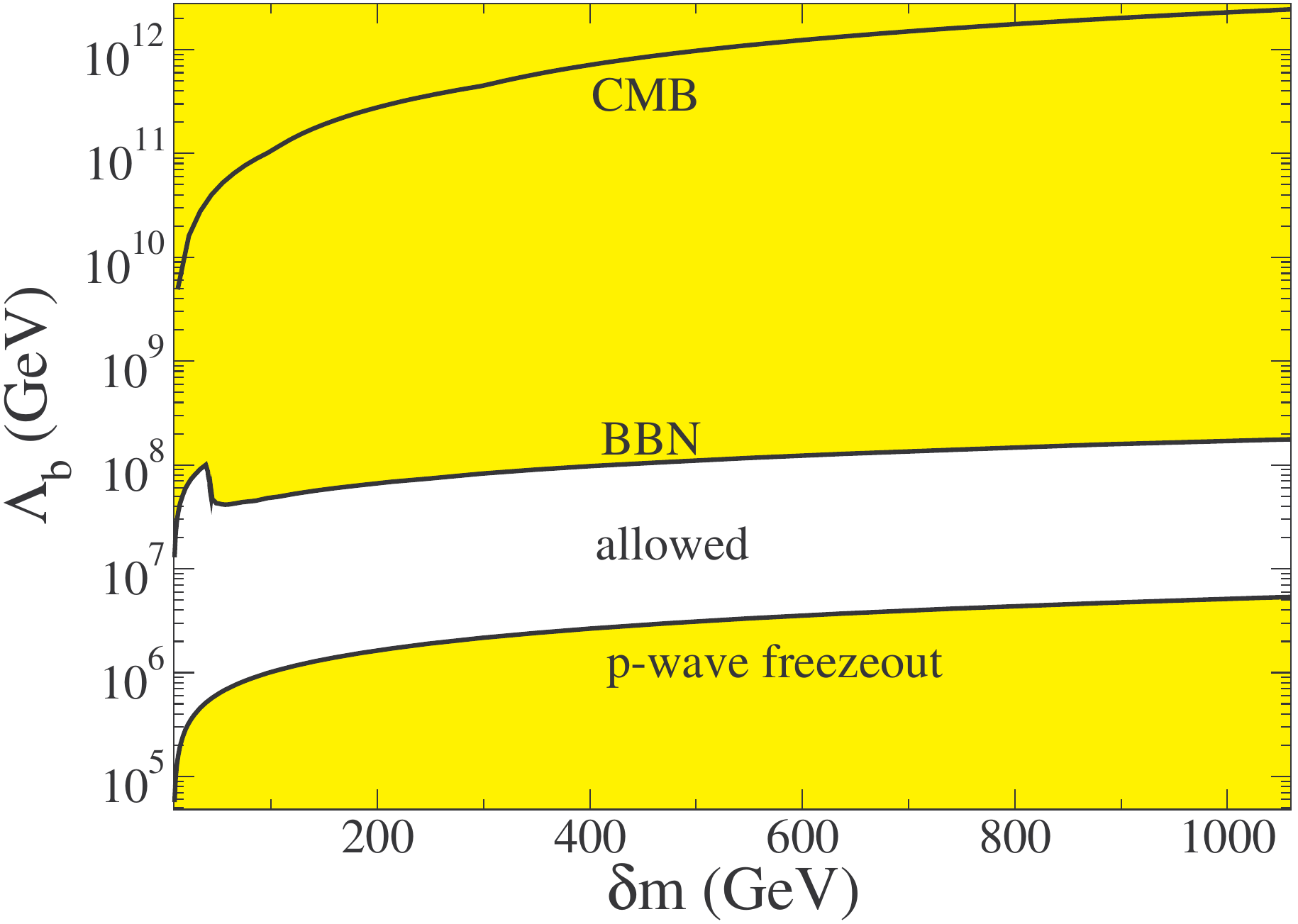}}
\caption{\label{Fig:Lifetime2} 
Similar to fig.\ \ref{Fig:Lifetime}, but for decays $\psi\to\chi b\bar b$. }
\end{center}
\end{figure}

\section{Constraints on decaying DM}
\label{constraints}

To ensure that $\psi$ decays occur after  freeze-out of $p$-wave annihilations
estimated in (\ref{xfeq}), we assume that $\Gamma < H(T_f)$ for the
relevant decay rate, with Hubble parameter $H(T_f) =
1.66\sqrt{g_*}(m_\chi/x_f)^2/M_p$ and $g_*\cong 76$ for $T_f\cong
3\,$GeV.  Comparing $H$ to the decay rates (\ref{rates}), we obtain
constraints on the  parameter space in figs.\
\ref{Fig:Lifetime},\ref{Fig:Lifetime2}, shown in the lower regions of
the plots. In the unshaded central regions, decays occur after freeze-out and
before big bang nucleosynthesis (BBN) or recombination.  In the upper
shaded regions, decays will disrupt BBN or the cosmic microwave
background (CMB), due to the deposition of electromagnetic energy, as
well as hadrons in the case of decays to $b\bar b$, as we consider in
the following subsections.

\subsection{BBN constraints}

For the first two operators of (\ref{eq:eff_op}), leading to decays into 
photons or electrons, only
the total energy deposited in the plasma is relevant for
photoproduction or dissociation of light elements produced by BBN.
We take the combined constraints
from ref.\ \cite{Cyburt:2002uv} (see fig.\ 8 of that reference).
An upper limit on $\zeta \equiv (n_\chi/n_\gamma)\,\delta m$ as a function
of lifetime is derived there, which we convert into a limit on
$\Lambda_{\gamma,e}$ as a function of $\delta m$, shown in fig.\
\ref{Fig:Lifetime} for $m_\chi=90\,$GeV. (The choice of $m_\chi$
determines $n_\chi/n_\gamma$.)  Since the limit
on $\zeta$ is not monotonic in lifetime, BBN excludes a range of
$\Lambda_{\gamma,e}$ for a given value of $\delta m$. 

The third operator of (\ref{eq:eff_op}) leading to $b\bar b$ pairs
entails somewhat more
stringent constraints because of hadronic interactions that can
more efficiently disturb light element abundances \cite{Jedamzik:2006xz}.  The limits depend not
only upon the total amount of energy deposited, but also the 
energy per decay.  By interpolating between the constraints of
\cite{Jedamzik:2006xz} calculated for different masses of decaying DM,
we find the BBN lower limit on $\Lambda_b$ versus $\delta m$ shown
in fig.\ \ref{Fig:Lifetime2}.\footnote{The relevant constraints are
inferred from figs.\ 9-10 of \cite{Jedamzik:2006xz}, in the region
$\tau < 100\,$s, which is insensitive to uncertainties in the
observed $^6$Li/$^7$Li abundance.}  The role of DM mass in that
reference (where the DM particle was assumed to decay completely to standard model particles) 
is played by $\delta m$ in the present context. 

\subsection{CMB constraints}
\label{cmb-constraints}

For lifetimes $\tau > 10^{12}\,$s, electromagnetic energy deposition 
starts to distort the cosmic microwave background, superseding 
BBN constraints.  We have computed the Planck-projected upper limits
on the injected energy fraction $\delta\Omega_\chi/\Omega_\chi = \delta
m/m_\chi$ as a function of
lifetime using the tools of ref.\ \cite{Slatyer:2012yq} (see also
ref.\ \cite{Cline:2013fm}), where
transfer functions $T_{\gamma,e}(z',z,E)$ are provided for computing the efficiency of energy
deposition as a function of redshift $z$ for injections of photons or
electrons at $z'$.  For $\chi\to\psi\,\gamma$, $T_{\gamma}$ can be used directly
since the spectrum is monochromatic.  For $\chi\to\psi\, e^+e^-$, 
$T_{e}$ must be convolved with the normalized energy spectrum of electrons from the
3-body decay, which in the limit of $\delta m\ll m_\chi$ takes the
form $dN/d\ln x = 60\, x^2(1-x)^3$, where $x=E/\delta m$.  Converting
the limits on $\delta m/m_\chi$ versus $\tau$ into the $\delta
m$-$\Lambda_e$ plane results in the excluded regions shown in fig.\
\ref{Fig:Lifetime}.  These extend to lifetimes greater than the age of
the universe, not of interest in the present context, since $\psi$ would still be the principal component
of the dark matter.

Projected Planck limits on the lifetimes for decays into $b\bar b$ have been given 
in ref.\ \cite{Cline:2013fm} for several DM masses.  Interpolating 
those results we translate them into 95\% C.L.\ limits on $\Lambda_b$ 
as a function of $\delta m$, shown in fig.\ \ref{Fig:Lifetime2}.

\begin{figure}[b]
\begin{center}
\centerline{
\includegraphics[scale=.33]{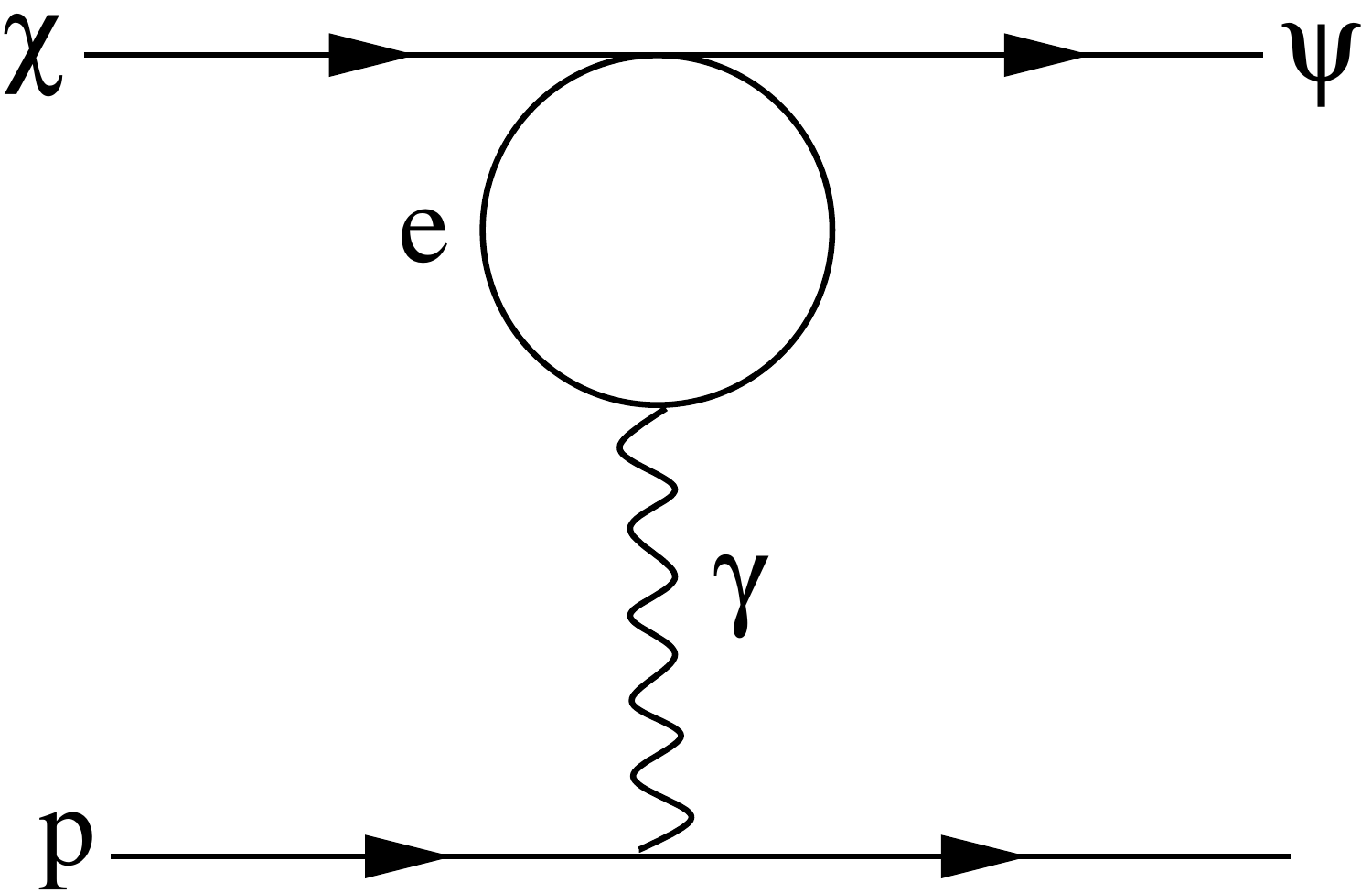}}
\caption{\label{Fig:loop} 
Loop-induced operator leading to inelastic scattering of DM on
protons. }
\end{center}
\end{figure}

\subsection{Direct detection}

For $\delta m\lesssim 100\,$keV, it is possible to have direct
detection through inelastic scattering on nuclei, $\chi N\to \psi N$.
This is relevant for the magnetic dipole operator for which such small
mass splittings are in the allowed region of fig.\ \ref{Fig:Lifetime}.
 We have roughly indicated the region
excluded by direct searches there by
taking the scattering rate to scale as $\Lambda_\gamma^{-2} f(v_{\rm
min}) \sim \Lambda_\gamma^{-2}\, e^{-3v^2_{\rm min}/2\sigma_v^2}$ where
$v_{\rm min}$ is the minimum velocity for an inelastic transition.
It is given in terms of the DM-nucleus reduced mass $\mu_{\chi N}$
as $v_{\rm min} = \sqrt{2\delta m/\mu_{\chi N}}$.  Therefore the
experimental limit on $\Lambda_\gamma$ takes the form
$\Lambda_\gamma > \Lambda_0\,e^{-\delta m/\delta m_0}$ for some reference
mass splitting, which we estimate to be $\delta m_0 \cong 6.3\,$keV
by comparison to recent constraints on magnetic inelastic dark matter
found in ref.\ \cite{Barello:2014uda}.  The coefficient $\Lambda_0$
corresponds to the limit from elastic scattering ($\delta m= 0$), which we take to be
$\Lambda_0\cong 10^{14}\,$GeV by rescaling the constraints on dipolar
dark matter from ref.\ \cite{Sigurdson:2004zp} according to the latest
limits from the LUX experiment \cite{Akerib:2015rjg}.

In principle, the four-fermion operators in (\ref{eq:eff_op}) could give
rise to inelastic scattering on nucleons, by forming a loop from the
electrons or quarks and considering virtual photon exchange between the 
loop and protons in the nucleus (see fig.\ \ref{Fig:loop}).  However the scattering rate is
negligible since the required mass splitting $\delta m > 2
m_e$ or $2 m_b$ is too large to be excited in direct detection experiments.
For smaller $\delta m$, there is an electron-loop mediated decay
$\psi\to\chi+3\gamma$ (decays into 1 and 2 photons are forbidden
by gauge invariance or Furry's theorem), but this is too slow to be
of interest for $\delta m < 2 m_e$, since the lifetime exceeds $10^{12}\,$s
and puts the model into the CMB-excluded region. 

\subsection{Fermi gamma ray line search}

The magnetic dipole operator in eq.\ (\ref{eq:eff_op}) induces $\chi
\bar \chi$ annihilation to monochromatic gamma rays through
$t$-channel exchange of a $\psi$ particle. The cross section for this
process has been calculated in  \cite{Weiner:2012cb}: 
\be
\langle \sigma v \rangle = \frac{16 m_\chi^4}{\pi \Lambda_\gamma^4} \left( \frac{m_\chi + \delta_m}{(m_\chi + \delta m)^2 + m_\chi^2} \right)^2 \, .
\ee

The Fermi-LAT collaboration has searched for such signals of DM
annihilation in the Milky Way halo \cite{Ackermann:2015lka}. In the
left plot in figure~\ref{Fig:Lifetime} we show the limits at 95\% C.L.
on the magnetic dipole operator from their search, assuming that the
DM density follows a generalized NFW profile with $\gamma = 1.2$
and corresponding to a region of interest of $3^{\circ}$ around the galactic center
to maximize the expected signal \cite{Ackermann:2015lka}. The line search limit is
$\Lambda_\gamma \gtrsim 8000$\ GeV, roughly equivalent to the bound we
obtained from coannihilations.

\subsection{Collider constraints}

The operators we consider are also constrained by collider searches.
For the $(\bar\chi\gamma^\mu\psi)(\bar e\gamma_\mu e)$ operator in 
eq.\ (\ref{eq:eff_op}), the relevant limits come from LEP, where the
characteristic signature is missing energy and a photon which is
radiated off the initial $e^+$ or $e^-$. For our fiducial case of
$m_\chi = 90$\ GeV, DELPHI monophoton searches constrain  $\Lambda_e
\gtrsim 310$\ GeV at 90\% C.L. \cite{Fox:2011fx}. For the magnetic
dipole operator, the current most stringent constraint is from LHC
monojet searches \cite{Barger:2012pf} requiring  $\Lambda_\gamma
\gtrsim 280$~GeV at 95\% C.L., a limit which is only slightly more
constraining than searches for monophotons at LEP \cite{Fortin:2011hv}
or the LHC \cite{Barger:2012pf}. The collider-disfavored region for the magnetic moment operator is more strongly excluded in our
scenario by direct detection and the $\psi$ lifetime constraints. The
$(\bar\chi\gamma^\mu\psi)(\bar b\gamma_\mu b)$ operator is 
in principle limited by LHC monojet searches, but the 
small $b$-quark content of the proton
makes such limits very weak.

All of the exclusions discussed here were derived under the assumption
that $e^{+} e^{-}$ or $p p$ collisions lead to stable  final-state
dark sector particles. Although it is possible for the $\psi$ produced
in these collisions to decay inside the detector, in the regions of
parameter space for which the collider limits are relevant, the mass
splitting is so small that the softness of the decay products would
render them undetectable.

For the operator of eq.\ (\ref{ann_op}), the relevant limits are again
from LEP monophoton searches when $f = e$ and LHC monojet searches
when $f=q$. Limits from an ATLAS monojet search
\cite{Aad:2015zva} as well as that from
the previously mentioned DELPHI monophoton search are shown in fig.\
\ref{fig:relic}. In either case the correct relic density is compatible with current collider limits.

\begin{figure*}[t]
\begin{center}
\centerline{
\includegraphics[scale=0.7]{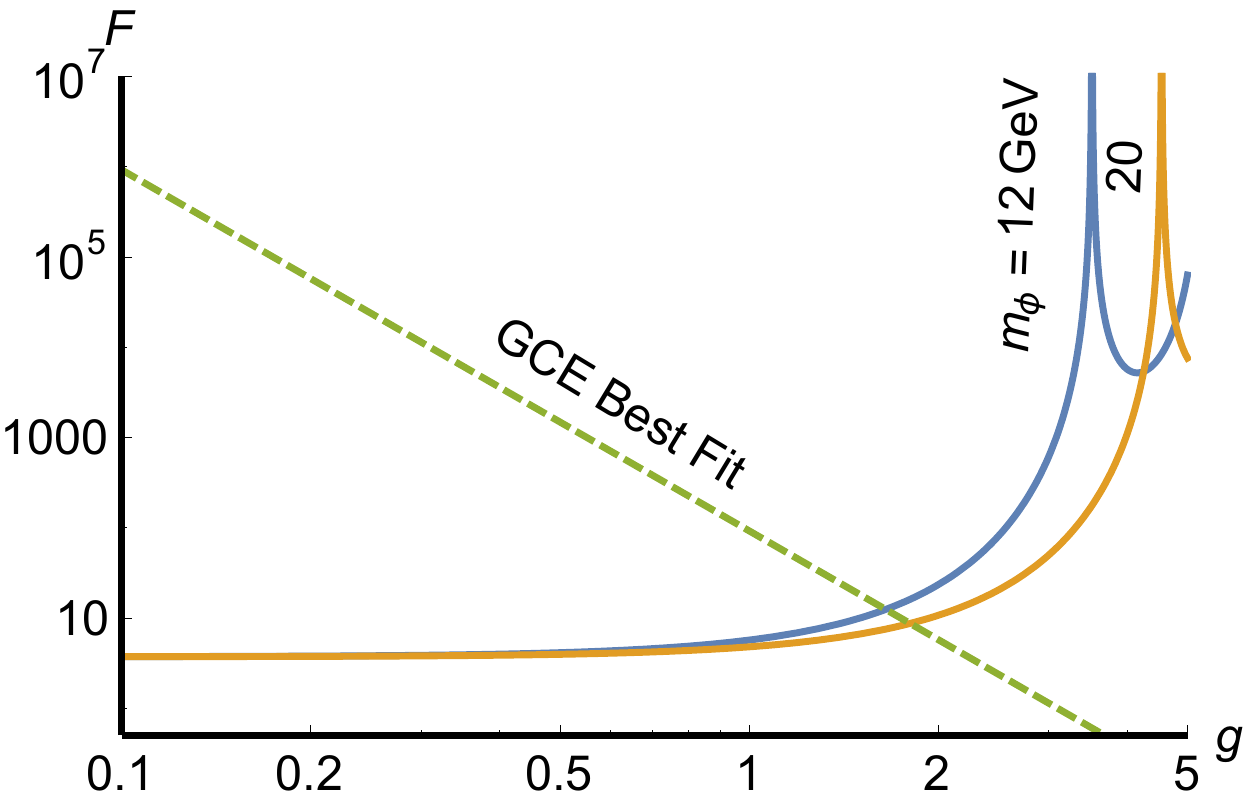}
\includegraphics[scale=0.7]{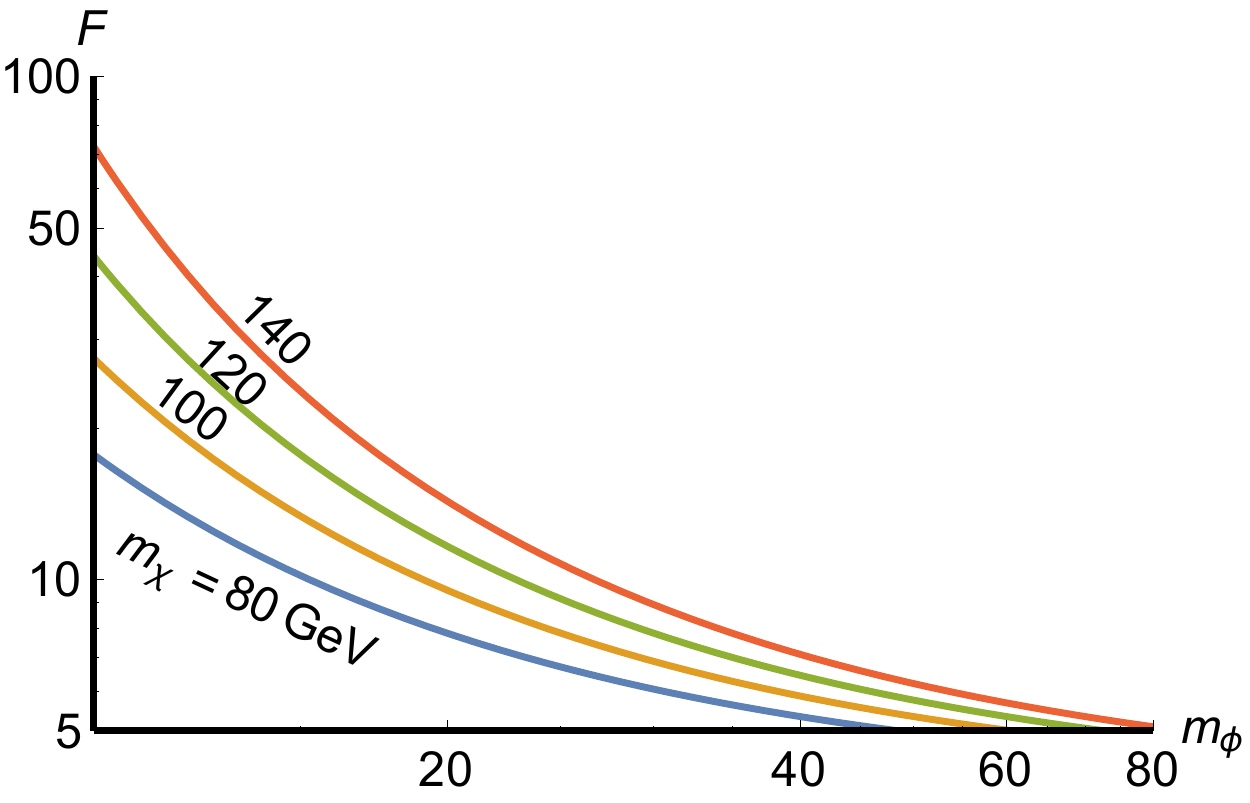}}
\caption{
Left: Phase-space averaged enhancement factor $F$ versus $g$
for $m_\chi=80\,\rm{GeV}$ and $\sigma_v=250\,\rm{km}\,\rm{s}^{-1}$, representative
of the Milky Way.  The corresponding result for 
$\sigma_v=10\,\rm{km}\,\rm{s}^{-1}$, appropriate for dwarf
spheroidals, looks very similar.
The solid lines show
$F(g)$ with for the cases $m_\phi=12\,\rm{GeV}$
and $m_\phi=20\,\rm{GeV}$. The dashed line
is the value of $F(g)$ required to give a sufficiently
large annihilation cross-section for the GCE. Right:
Dependence of $F$ on $m_\phi$ for a fixed value of $g=0.8$ (the
intersection point in the top left panel) with masses
$m_\chi=80,\,100,\,120,\,\rm{and}\,140\,\rm{GeV}$.}
\label{Fig:sommerfeld}
\end{center}
\end{figure*}

\section{Mediator couplings}
\label{mediator}

A large $p$-wave cross section would generically run afoul of 
direct and indirect detection constraints if the DM $\chi$ coupled
directly to SM particles. On the other hand,
annihilation to
light mediators $\phi$, that subsequently decay into SM particles, 
can avoid this problem.  If $\chi$ couples to $\phi$
as $g\phi\bar\chi\chi$, the resulting $p$-wave cross section at low velocities is given
by
\be
	\sigma v \cong {3 v^2 g^4\over 32\pi m_\chi^2}
	\label{eq:cs_on-shell}
\ee
An uncomfortably large coupling $g\sim 3.7$ would be needed to match the fit to the GC
excess. 

However smaller values of $g$ can be sufficient if the cross section
is Sommerfeld enhanced, which can naturally occur if the mediator 
$\phi$ is light.  Defining $\alpha_g = g^2/4\pi$, 
 an analytic approximation to the enhancement
factor is given by \cite{Cassel:2009wt}
\begin{align}
	S_l &\cong \left|\frac{\Gamma(a^+)\Gamma(a^-)}{\Gamma(1+l+2iw)}\right|^2.
\label{eq:Sommerfeld}
\end{align}
for partial wave $l$ scattering, 
where 
\bea
a^\pm&=&1 + l+iw\,\left(1\pm\sqrt{1-x/w}\right)\nonumber\\
x&=&{\alpha_g\over \beta},\quad 
w=\frac{6\,\beta\, m_\chi}{\pi^2\,m_\phi}
\eea
with velocity $v = \beta c$ in the center of mass frame.
For a $p$-wave process we take $l=1$. $S_1$ is nonvanishing in the limit $v\to 0$, so that 
velocity suppression of the $p$-wave cross section is still present
despite the enhancement, and $S_1$ has
quasiperiodic resonant behavior as a function of $\alpha_g$.

The enhancement factor depends on the relative velocity of the 
particles, which in principle must be averaged over phase space. 
Ignoring the radial dependence of the annihilation cross-section, 
we can find an estimate of the average enhancement, which is given
 by
\be
F(g)= {\langle S_1\, \sigma v_{\rm rel}\rangle\over
	\langle \sigma v_{\rm rel}\rangle}
\ee
using eq.\ (\ref{eq:velavg}). 
However for the parameter values of interest, we find that the
dependence on $v$ is very weak and one can simply use $\sim 10$-$1000$
km/s with negligible error from omitting the average. 
To match the desired value of the cross
section in (\ref{eq:fit}), we need
\be
	g^4 F(g) = 144\left(m_\chi\over 100{\rm\,GeV}\right)^2
\ee
This relation as a function of $g$ is shown as the dashed line in fig.\ 
\ref{Fig:sommerfeld}, while the analytic approximation for $F$ is
indicated by the curves for two different values of the mediator mass.

By comparing $F(g)$ to the  required cross section,
eq.~(\ref{eq:fit}), we find that the coupling constant can be reduced
to a more comfortable value of $g\cong 2$. It can be somewhat further reduced
by taking larger values of $m_\chi/m_\phi$, as can be seen in
figure~\ref{Fig:sommerfeld}; the left panel shows $F(g)$ decreasing
with $m_\phi$, while the right shows $F(g)$ increasing with $m_\chi$.
It was recently pointed out \cite{Blum:2016nrz} that approximations
to the enhancement factor such as (\ref{eq:Sommerfeld}) may fail to
satisfy partial wave unitarity in the resonant regions.  We have
checked that we are very far from any such violation however, for the 
parameter values of interest.

Finally, it has been noted in \cite{An:2016kie} that it is possible for two DM particles to capture into a bound state and then annihilate to mediators. The bound state formation process dominantly occurs in the $s$-wave, so, if possible, it dominates over the direct $p$-wave annihilation to mediators. In forming a bound state a mediator is emitted, so the mass of the mediator must be less than the binding energy of the ground state for this to occur, i.e.
\begin{equation}
m_\phi \leq \frac{g^4 m_\chi}{64 \pi^2} \, .
\end{equation}
For the values of $g$ and $m_\chi$ that we consider to explain the GC excess, $m_\phi \lesssim 2.2$ GeV for a bound state to form. In this work we have only considered mediator masses above this limit.

\section{Conclusion}
\label{conclusion}

We have presented a scenario in which $p$-wave annihilating dark
matter could have significant indirect signals from the galactic
center despite having a velocity-suppressed cross section.  Although
our immediate motivation was to reconcile a dark matter interpretation
of the observed GC gamma ray excess with conflicting constraints from
dwarf spheroidal galaxies, the framework presented here could be of more
general interest.  

Our key idea is to assume that the current generation of $p$-wave
annihilating DM $\chi$ is the decay product of a metastable
predecessor particle $\psi$, which has a thermal origin and decays
after $p$-wave annihilations of the stable DM have frozen out.  This
allows a large range of lifetimes that depend upon the $\psi$-$\chi$
mass splitting $\delta m$ and the mass scale $\Lambda$ of the
effective interaction which leads to the decay.  A number of
constraints must be satisfied, including those coming from  BBN, the
CMB, direct detection, photon line searches, mono-X searches at
colliders, and $\psi f\to\chi f$ scattering in the early universe
(which could deplete the DM abundance).  They depend strongly upon the
nature of the decays, which we have illustrated using the examples of
$\psi\to \chi e^+ e^-$, $\psi\to \chi\gamma$, and $\psi\to\chi b\bar
b$, but in all cases there is a significant region of the $\delta
m$-$\Lambda$ parameter space in which all of the requirements can be
satisfied.

The annihilation cross sections of interest for explaining the galactic center
excess are larger than would generically occur in the presence of
$p$-wave suppression because of the low DM velocity in the GC.  We
nevertheless demonstrated a working example using reasonable coupling
strengths, where the DM annihilates into light scalar mediators that mix
with the Higgs boson and subsequently decay into $b$ quarks.  We have
shown that such models give a reasonably good fit to the observed GCE,
while satisfying constraints from dwarf spheroidals by a 
comfortable margin.  

Collider tests of our scenario are currently weaker than the 
consistency requirement that inelastic $\psi f\to\chi f$ scatterings
on standard model particles do not deplete the DM relic density in the
early universe (due to strong $p$-wave annihilations of $\chi$).
For a narrow window of mass splittings $\delta m\sim 0.1\,$MeV,
direct SM searches provide a possible means of detection in the case
of magnetic inelastic transitions. 

 Fermi observations of gamma rays from galaxy clusters may provide a
more sensitive test of our scenario, due to the  large velocity
dispersion in clusters. We have shown that limits on DM annihilation
from the Virgo cluster, while significantly stronger than limits from
dwarf spheroidals, still are far from being in tension with this
interpretation of the GCE. We hope that our work will motivate further
studies of limits on DM annihilation in the Coma cluster, which 
has the potential to be more constraining because of its relatively
high velocity dispersion.

\medskip
{\bf Acknowledgments}.  
We thank Kimmo Kainulainen, Maxim Pospelov, Subir Sarkar, Tracy Slatyer, Alex Drlica-Wagner and 
Wei Xue for helpful discussions or correspondence. We acknowledge support of the
McGill Space Institute and the Natural Sciences and Engineering
Research Council of Canada.  We thank Niels Bohr International Academy
for its generous hospitality during the initial stages of this work.

\bibliography{pwavebib}{}
\bibliographystyle{hep}

\end{document}